\documentclass[%
 reprint,
superscriptaddress,
 amsmath,amssymb,
 aps,
 pre,
]{revtex4-1}

\usepackage{graphicx}
\usepackage{dcolumn}
\usepackage{bm}
\usepackage{xcolor}
\usepackage[bf]{subfigure}

\begin{document}

\title{Modeling how social network algorithms can influence opinion polarization}

\author{Henrique Ferraz~de~Arruda}
\affiliation{S\~ao Carlos Institute of Physics,
University of S\~ao Paulo, S\~ao Carlos, SP, Brazil}
\affiliation{Institute for Biocomputation and Physics of Complex Systems (BIFI), University of Zaragoza, Zaragoza, Spain}

\author{Felipe Maciel~Cardoso}
\affiliation{Institute for Biocomputation and Physics of Complex Systems (BIFI), University of Zaragoza, Zaragoza, Spain}

\author{Guilherme Ferraz~de~Arruda}
\affiliation{ISI Foundation, Turin, Italy}

\author{Alexis R. Hern\'andez}
\affiliation{Institute of Physics, Rio de Janeiro Federal University, Rio de Janeiro, RJ, Brazil}

\author{Luciano da~Fontoura~Costa}
\affiliation{S\~ao Carlos Institute of Physics,
University of S\~ao Paulo, S\~ao Carlos, SP, Brazil}

\author{Yamir Moreno}
\affiliation{Institute for Biocomputation and Physics of Complex Systems (BIFI), University of Zaragoza, Zaragoza, Spain}
\affiliation{ISI Foundation, Turin, Italy}
\affiliation{Department of Theoretical Physics, Faculty of Sciences, University of Zaragoza, Zaragoza, Spain}

\begin{abstract}
Among different aspects of social networks, dynamics have been proposed to simulate how opinions can be transmitted. In this study, we propose a model that simulates the communication in an online social network, in which the posts are created from external information. We considered the nodes and edges of a network as users and their friendship, respectively. A real number is associated with each user representing its opinion. The dynamics starts with a user that has contact with a random opinion, and, according to a given probability function, this individual can post this opinion. This step is henceforth called \emph{post transmission}. In the next step, called \emph{post distribution}, another probability function is employed to select the user's friends that could see the post. Post transmission and distribution represent the user and the social network algorithm, respectively. If an individual has contact with a post, its opinion can be attracted or repulsed. Furthermore, individuals that are repulsed can change their friendship through a rewiring. These steps are executed various times until the dynamics converge. Several impressive results were obtained, which include the formation of scenarios of polarization and consensus of opinions. In the case of echo chambers, the possibility of rewiring probability is found to be decisive. However, for particular network topologies, with a well-defined community structure, this effect can also happen. All in all, the results indicate that the post distribution strategy is crucial to mitigate or promote polarization.
\end{abstract}

\maketitle

\section{Introduction}
With the advent of the internet, many different online social networks have been created. In order to understand the impact of these networks on the users' opinions, different models have been proposed~\cite{castellano2009statistical, noorazar2020recent,urena2019review}. The simulated dynamics include voter model~\cite{garcia2014voter}, majority rule model~\cite{galam2002minority}, bounded confidence model~\cite{lorenz2007continuous}, among others~\cite{castellano2009statistical}. Part of these studies considers the dynamics executed on a network structure, in which the nodes and edges represent people and their friendship, respectively. Several distinct characteristics of social networks have been studied in order to understand opinion dynamics. For example, \emph{Sznajd et al.}~\cite{sznajd2000opinion} considered that when two or more people have the same opinion, it is more likely for them to convince others.  In general, these studies consider static network structures. However, other researchers take into account that people can change their friendship according to time~\cite{durrett2012graph,fu2008coevolutionary,gracia2011coevolutionary,he2004sznajd,holme2006nonequilibrium,iniguez2009opinion,su2020noise}. In this case, edge rewirings are employed, giving rise to groups of connected people with similar opinions, called echo chambers.

One essential characteristic of these dynamics is how to represent the opinion. In many cases, the opinions are expressed only for two possible states~\cite{garcia2014voter,galam2002minority,sznajd2000opinion}. In other cases, a varied number of categories~\cite{de2019contrarian,holme2006nonequilibrium}, or vectors~\cite{baumann2020emergence} can also describe the opinions. Another option is to express opinions as a continuous number~\cite{baumann2020modeling,deffuant2000mixing,lorenz2007continuous,oestereich2020hysteresis,queiros2016interplay,sasahara2019inevitability}, which can express problems regarding negotiations. In this case, opinions are not categorical, and the individuals can have intermediate opinions. Promising results have been obtained from this type of dynamics. For instance, in~\cite{baumann2020modeling}, results obtained from simulations were found to be similar to the scenario observed in the online social networks. 

Although in real social networks, people typically have lots of friends, in~\cite{de2019contrarian,benatti2020opinion}, the authors considered that a person is not capable of interacting with lots of people. For this reason, they adopt network models with low average degrees. In order to reduce the interactions between individuals, we considered two complementary mechanisms. The first represents the user's action of posting pieces of information, henceforth called \emph{post transmission}. In contrast with other approaches, the individual can post something different from their own opinion. In the following, we simulate the individuals' information transference, a mechanism that chooses if the data should be delivered to other users. We named this mechanism as \emph{post distribution}. This step simulates the possibilities of how social network algorithms can manage posts. According to the received posts, opinions can change positively or negatively, which we henceforth called \emph{attraction} and \emph{repulsion}, respectively. Furthermore, the individual can rewire the respective connection for the cases where the post repulses the individual's opinion.

In order to model post transmission, we compare functions that represent different scenarios. The first consists of users who post information they like or dislike, which can be understood as a reaction in a social network.  We also analyzed users that only post information they like. In the third case, we considered users that did not pay attention to their posts. As the latter, we take into account users with varied behaviors. In the case of post distribution, we also test many distinct options as this can play an essential role in social networks to hold the attention of the users. 

We analyzed the obtained results by employing two distinct and complementary types of measurements—the first consists of analyzing the resulting opinion distributions. We quantify how polarized and balanced the opinion distributions are. However, more information can be obtained if the network structure is considered.  As well as in~\cite{baumann2020modeling,cinelli2020echo,cota2019quantifying}, we measured the relationship between the node opinion and the average of the friends' opinions. From this analysis, different resulting structures can be observed, which include the presence of echo chambers. Our model can give rise to echo-chambers for a specific scenario, even without friendship rewirings. Additionally, the bimodality of the opinion distribution does not guarantee that the dynamics would converge to echo-chambers. We also compared our dynamics with networks obtained from Twitter. Interestingly, we found similarities between the obtained results with the proposed dynamics, including the level of bimodality and echo chambers.

The paper is organized as follows. In section~\ref{sec:opinion_dynamics}, we present our proposed dynamics, as well as the measurements employed to analyze the results. Section~\ref{sec:reults} describes the obtained results and their respective discussion. Finally, in Section~\ref{sec:concusions}, we conclude the paper and present future works' perspectives.

\section{Opinion dynamics}
\label{sec:opinion_dynamics}
This section describes the proposed dynamics and depicts the experiment design and how the results are analyzed.

\subsection{Proposed framework}

\begin{figure*}[!htpb]
    \centering
     \includegraphics[width=1.\textwidth]{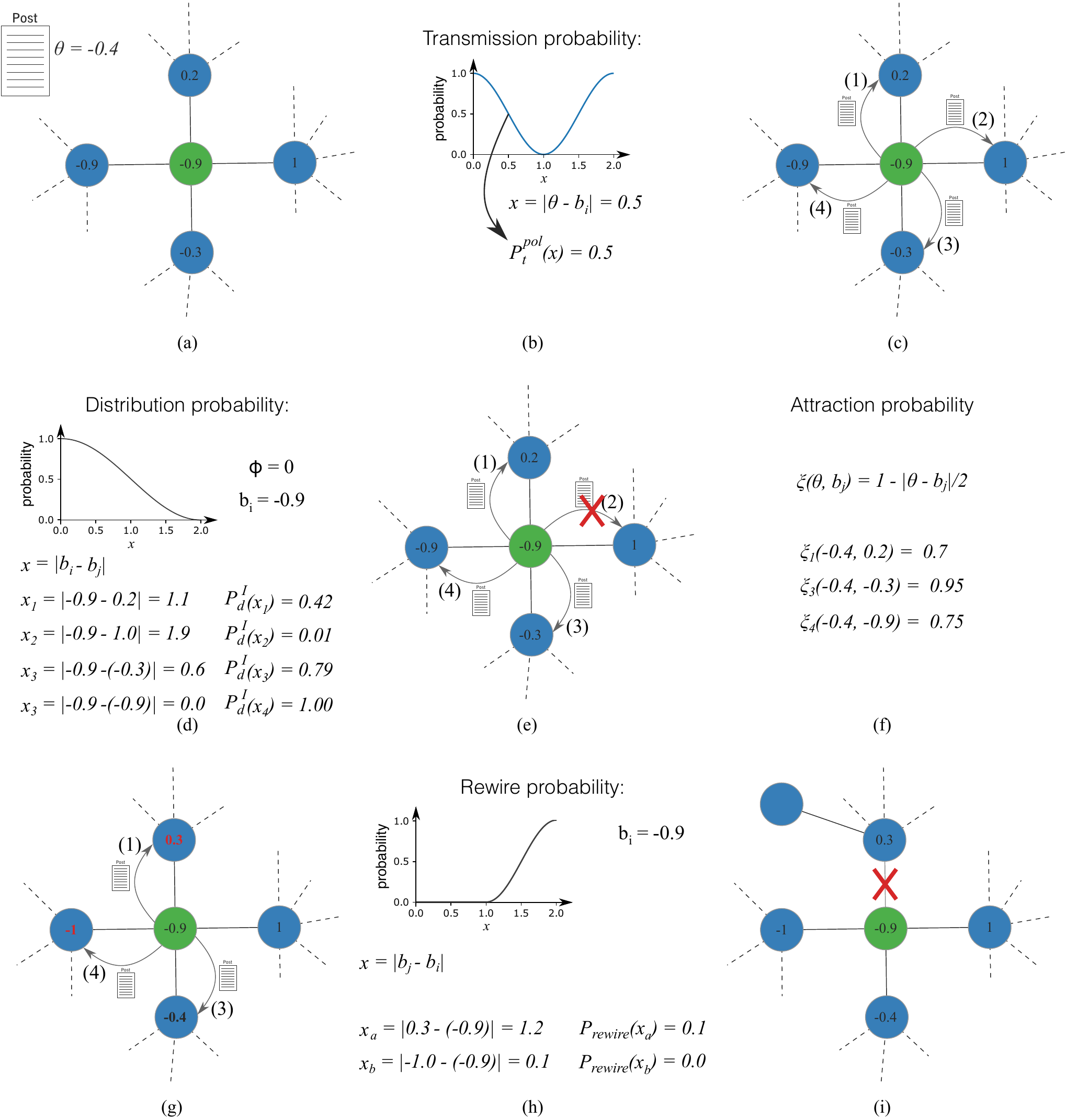}
   \caption{Example of one step of the proposed dynamics. The number attributed to each node represents its opinion. First, a node (green) is selected at random. In this case, the opinion of the chosen individual is $b_i=-0.9$. In (a), a number ($-1 \leq \theta \leq 1$) is randomly generated according to a uniform distribution, representing a post created by the green individual. The post is given by $\theta= -0.4$. In (b), the transmission probability, $P_t(x)$, is calculated according to the difference between the post and the individual's opinion. We illustrate our example by employing the polarized function $P_t^{\text{pol}}$. If the post is transmitted, in (c), it can be distributed. (d) illustrates how the probabilities of \emph{post distribution} are calculated, which are exemplified by using $P_d^{\text{II}}$ (see Section~\ref{sec:post_distributions_pos}). According to these probabilities, in (e), the algorithm chooses if the post is seen by the other users (blue). We calculate the attraction probability for all individuals who receive the post, as shown in (f). As a complement, the non-attracted individuals are repulsed. (g) illustrates the opinion changes, represented in bold. The rewiring probability is computed for all repulsed users, as shown in (h), and, in (i), the edges can be rewired. The new connection is chosen with the same probability of reconnection for all remaining network nodes.}
  \label{fig:scheme}
\end{figure*}

Our model is based on a social network, in which the users (individuals) produce a \emph{posts}. In the following, the social network algorithm (post distribution) selects the neighbors (friends) that will receive the post. Finally, the opinions of the selected neighbors are changed, and the users that strongly disagree can change their friendship (rewiring). Figure~\ref{fig:scheme} illustrates one iteration of the proposed dynamics.

Each node, $i$, represents an individual that stores an opinion, $b_i$, as a real number, in which is $-1 \leq b_i \leq 1$. The network edges represent the individual's friendships. The dynamics start with the opinions randomly initialized by following uniform probability distribution.

For each iteration, a node $i$ is randomly selected, and a post $\mathcal{P}$ is created (see Figure~\ref{fig:scheme}~(a)). In order to define the post's opinion, a number, $\theta$, is randomly generated with uniform probability ($-1 \leq \theta \leq 1$). In the following, a transmission probability, $P_t(x)$, is used to define if the individual $i$ will post $\mathcal{P}$. This probability function is computed according to the difference between the post and the individual's opinion ($x = |\theta - b_i|$), as shown in Figure~\ref{fig:scheme}~(b). The functions employed in this paper are described in Section~\ref{sec:prob_func}.

In the following, if the individual $i$ produces a post (Figure~\ref{fig:scheme}~(c)), there is a probability function, $P_d(x)$, defined for all of the $i$ neighbors to receive the information. In this case, the function is calculated for all edges, $(i,j)$, connected to $i$, where $x=|b_i - b_j|$. One example of this action is shown in Figure~\ref{fig:scheme}~(d)~and~(e). This action is associated with how the social network algorithm acts. The used probability functions are shown in Section~\ref{sec:prob_func}. After this action, another probability could be associated with the individuals to define if they are active in the social network. However, here we consider this probability as one. In other words, we considered that the users take a look at all received posts.

The opinions of the individuals that receive the post can be attracted or repulsed. More specifically, for each individual, $j$, the probability of being attracted is 
\begin{equation}
    \xi_j(\theta, b_j) = 1 - \frac{|\theta - b_j|}{2}.
\end{equation}
An example of this step can be seen in Figure~\ref{fig:scheme}~(f). 

As observed in \cite{Isenberg1986,Moscovici1969}, people update their opinions on subjects after interacting, or in a discussion, and can become more polarized while doing so. Thus, in our model, if the individual is attracted, its opinion $b_j$ turns to be $b_j + \Delta$, where $\Delta$ is a real number. In the cases in which $\theta$ is lower than $b_j$, $\Delta$ becomes a negative number, otherwise positive. It has also been observed that, when confronted with opposing views, people in social media can become more extreme in their opinions~\cite{Bail2018}. Thus, to incorporates these effects in our dynamics, if not attracted, the individuals are repulsed, where $b_j$ turns to be $b_j - \Delta$. Furthermore, if the resulting $b_j<-1$ or $b_j>1$, $b_j$ turns to be $-1$ or $1$, respectively, which is illustrated in Figure~\ref{fig:scheme}~(g). 

High values of $\Delta$ strongly affect the dynamics because it leads the distributions to be less well-defined. Furthermore, with low values, the dynamics delays much more in converging. So, for all of our analyses, we empirically adopted $|\Delta|=0.1$. 

As the last action of our dynamics, we allow an individual to unfollow a given friend and connects to another at random (see Figures~\ref{fig:scheme}~(h)~and~(i)). This step is henceforth called \emph{rewire}. The unfollowing in social networks have been extensively studied~\cite{kwak2011fragile,kwak2012more,xu2013structures}. For instance, the study developed by~\cite{kwak2012more} indicates that Twitter users are less likely to unfollow friends who have acknowledged them. With the basis in this study, here, if an individual is repulsed by a neighbor, the rewire can happen according to a given function, $P_{\text{rewire}}(x)$, for $x = |b_i - b_j|$. Our employed strategies are described in Section~\ref{sec:prob_func}.

Another initial node is randomly selected, and all the process is repeated $n$ iterations, in which $n$ should be big enough to lead the dynamics to reach a steady state. In order to automatically execute many times the same program with all of the parameters presented in this section, we use the software \emph{GNU Parallel}~\cite{tange2011gnu}.

\subsection{Adopted configurations}
\label{sec:prob_func}
In this subsection, we describe the adopted possibilities of the probability functions and the respective motivations. 

\subsubsection{Post transmission functions}
\label{sec:post_transm_pos}
In the case of post transmission, we considered three distinct possibilities of $P_t(x)$. The first possibility is defined as
\begin{equation}
    P_t^{\text{pol}}(x) = \cos^2\left(x \frac{\pi}{2}\right),
\end{equation}
where $x = |\theta - b_i|$.
In this case, the individual tends to post both the most similar and most different subjects than its own belief. This polarized function simulates the scenarios in which the users post pieces of information he/she agrees or disagrees. In the latter, this possibility represents the cases in which a user post reflects an opinion against the content. Furthermore, the highest probabilities of posting divergent opinions are reached only if the individuals' opinions, $b_i$, are close to the extremes, $-1$ or $1$, resulting from the maximum possible value of $x$. For instance, if $b_i=0$, the maximum $x$ value is 1, consequently, $P(1) = 0$.

We also considered the users that have a much higher probability of posting information similar to their own opinions and cannot post contrarian information, which can be modelled as 
\begin{equation}
    P^{\text{sim}}_t(x)= 
\begin{cases}
    \cos^2\left(x \frac{\pi}{2}\right), & \text{if } x\leq 1\\
    0,              & \text{otherwise}.
\end{cases}
\end{equation}

The third tested strategy is the uniform probability, as follows  
\begin{equation}
    P^{\text{uni}}_t(x) = 1.
\end{equation}
This probability simulates the cases in which the users produce posts without taking care of the information. More specifically, all the created posts are spread by the users. This case can also be used as a null model.

In addition to the three described possibilities, we considered the combination among them, $P^{\text{all}}_t(x)$. A function is randomly chosen for each individual, with equal probability for all possibilities. For the sake of simplicity, each individual has a fixed behavior. More specifically, the chosen function does not change during the dynamics execution. 

\subsubsection{Post distributions}
\label{sec:post_distributions_pos}
\begin{figure}[t]
  \centering
    \subfigure[$P_d^I(x)$]{\includegraphics[width=0.24\textwidth]{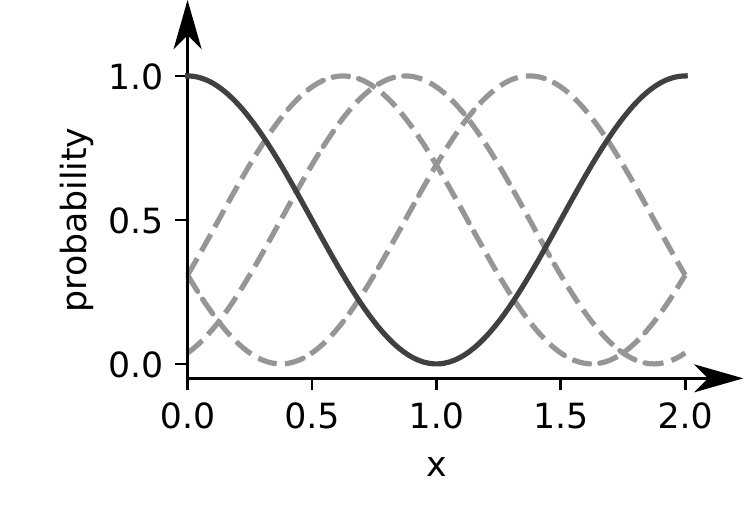}}
    \subfigure[$P_d^{II}(x)$]{\includegraphics[width=0.24\textwidth]{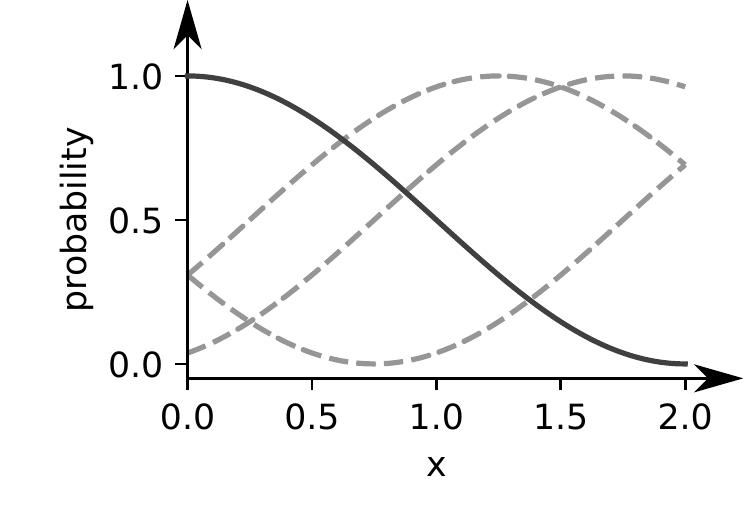}}
    
  \caption{Examples of distribution probability functions. The black curves represent the probabilities for $\phi=0$.}
  \label{fig:examples_cos}
\end{figure}

In the case of post distribution, we also took into account some possibilities of probability functions. The first equation is defined as follows
\begin{equation}
    P_d^{\text{I}}(x) = \cos^2\left(x \frac{\pi}{2} + \phi \right),
    \label{eq:rp_I}
\end{equation}
where the parameter $\phi$ is a real number that controls the starting point of the cosine-squared function and $x = |b_i - b_j|$\footnote{This process is similar to a homophily, however, homophily is an outcome of people decisions, while here we are modeling the posts suggestion by the social network algorithm. Homophily could be an extension to our model.}, in which $b_i$ and $b_j$ are the opinions of the individual $i$ and its given neighbor $j$, respectively. We considered another version, in which the probabilities varies smoother than in equation~(\ref{eq:rp_I}), as follows 
\begin{equation}
    P_d^{\text{II}}(x) = \cos^2\left(\frac{x}{2} \frac{\pi}{2} + \phi \right).
\end{equation}
Figure~\ref{fig:examples_cos} illustrates the relationship between the parameter $\phi$ and the functions of probability. For both cases, the functions can represent a range of algorithms that spread information from a polarized to depolarized. As the third case, we also employed a null model, that transmits uniformly the information, which is defined as 
\begin{equation}
    P_d^{\text{III}}(x) = 1.
\end{equation}

\subsubsection{Rewiring configurations}

In order to rewire only connections between individuals that strongly disagree, we adopt the following rewiring probability function
\begin{equation}
    P_{\text{rewire}}(x)= 
\begin{cases}
    \cos^2\left(x \frac{\pi}{2}\right), & \text{if } x > 1\\
    0,              & \text{otherwise},
\end{cases}
\label{eq:rewiring}
\end{equation}
where $x$ is defined by the difference between the opinions of the individuals $i$ and $j$ ($x = |b_j - b_i|$). We also considered the dynamics without the possibility of rewiring ($P_{\text{rewire}}(x) = 0$). 

\subsection{Opinion polarization analysis}

In this section, we describe how the results are analyzed.  First, we present the employed measurements to account for the opinion distributions. Next, we also considered information regarding the relationship between topology and the network structure. 

\subsubsection{Analysis of opinion distributions}

By employing the bimodality coefficient, $BC$~\cite{pfister2013good}, we can quantify the level of polarization of the opinions. This measurement is computed from the opinion distributions and is defined as
\begin{equation}
    BC = \frac{g^2 + 1}{k + \frac{3(n-1)^2}{(n-2)(n-3)}},
\end{equation}
where $n$ is the number of samples, and $g$ and $k$ are the \emph{skewness}~\cite{kokoska2000crc} and \emph{kurtosis}~\cite{decarlo1997meaning} of the analyzed distribution, respectively. Furthermore, it was empirically found that for $BC_{\text{critic}} = 5/9$ the distribution tends to be uniform, and for values higher and lower then $BC_{\text{critic}}$, it tends to be bi-modal and uni-modal, respectively~\cite{pfister2013good}. 
However, in our experiments, $BC$ does not performed well in probability distributions with unbalanced modes. In order to complement the understanding of this measurement, we propose a measurement that is henceforth called \emph{balance}. First, we divide the resulting opinions into two sets, $s_1$ and $s_2$, which contains values lower and greater than zero, respectively. From these sets, we compute the balance, as follows
\begin{equation}
\beta = \frac{\min{(c_1,c_2)}}{\max{(c_1,c_2)}},
\end{equation}
where $c_1$ and $c_2$ represent the number of samples in $s_1$ and $s_2$, respectively.

\subsubsection{Relationship between topology and dynamics}
\begin{figure*}[!htpb]
  \centering
    \subfigure[Consensus
    ]{\includegraphics[width=0.24\textwidth]{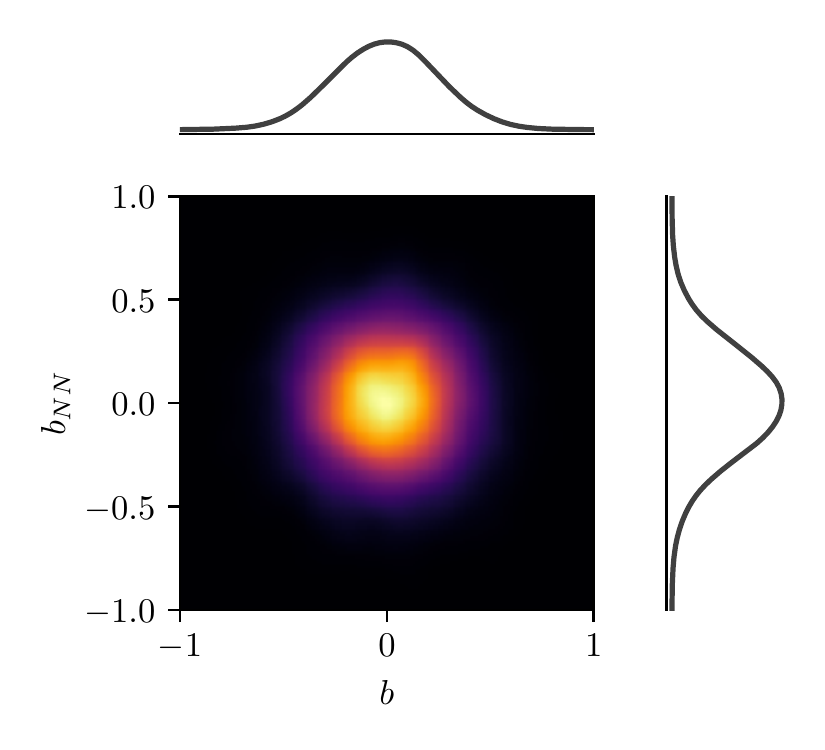}}
    \subfigure[Echo chamber,
    ]{\includegraphics[width=0.24\textwidth]{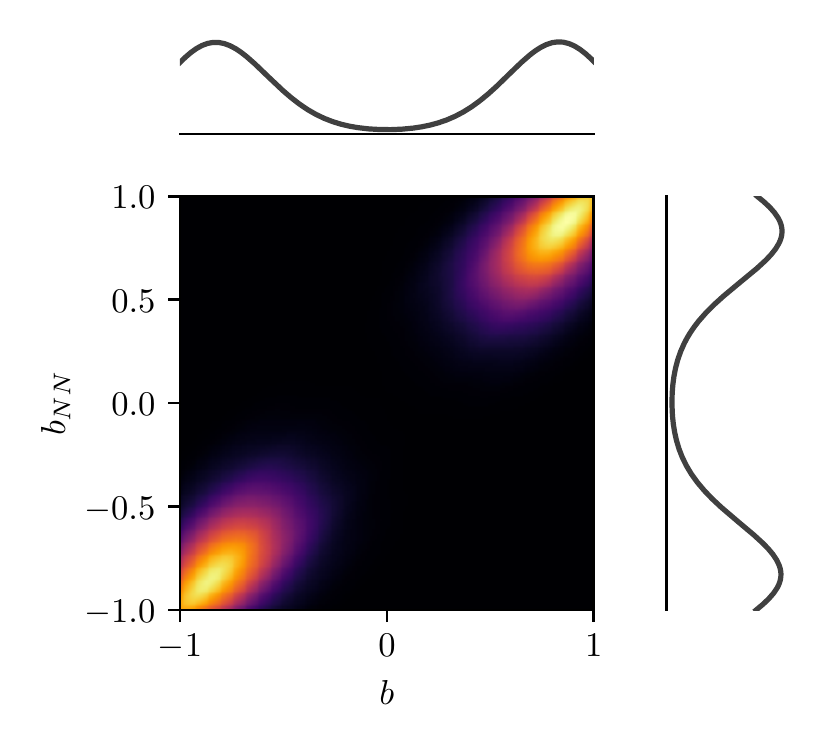}}
    \subfigure[Diverse
    ]{\includegraphics[width=0.24\textwidth]{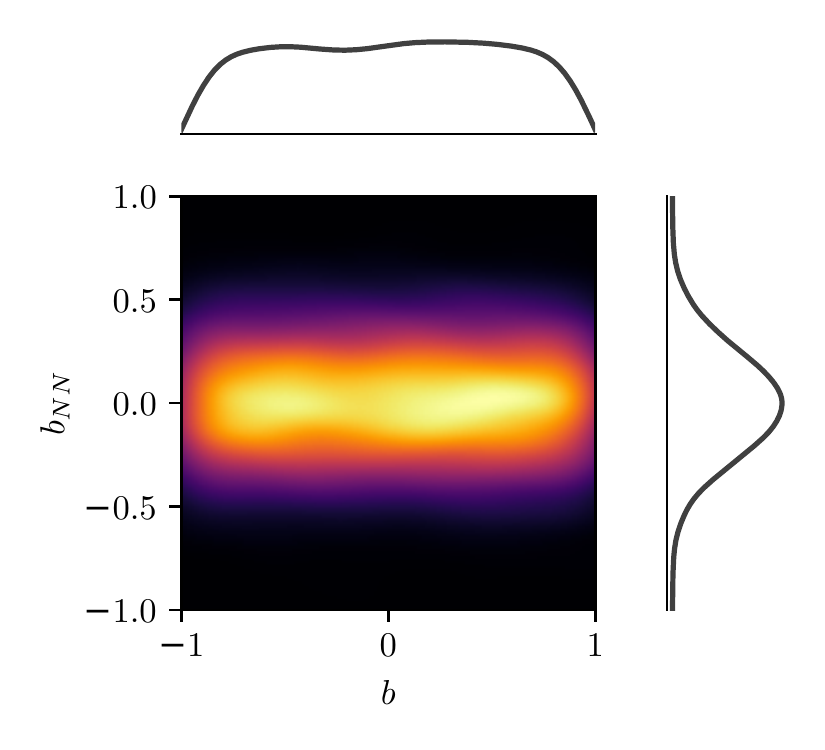}}
    
  \caption{Artificial examples of possible resulting density maps of opinions $b$ against the average neighbor's opinions $b_{NN}$, in which the lighter color represent the larger number of users. Furthermore, these distributions depict the probability functions of $b$ and $b_{NN}$.}
  \label{fig:examples_pol}
\end{figure*}

Although the bimodality coefficient accounts for the opinion distribution's shape, it cannot quantify individuals' relationships. So, we employ another analysis. An interesting characteristic that can be found in opinion dynamics is the presence of echo chambers. Here, we consider the measurement used in~\cite{baumann2020modeling,cinelli2020echo,cota2019quantifying} to identify if our dynamics leads to the formation of echo chambers, which consists of a density map of the individuals' opinion, $b$, against the average opinion of its neighbors, $b^{NN}$. So, when distinct groups are located in the first and third quadrants of the map, the dynamics converged to echo chambers. However, other interpretations can be taken. In order to illustrate some possibilities of resulting density maps, we create three case examples (see Figure~\ref{fig:examples_pol}). In Figure~\ref{fig:examples_pol}~(a), a single peak expresses \emph{consensus}, in which all individuals are connected to others with similar opinions. Another possible scenario can also be depicted in Figure~\ref{fig:examples_pol}~(c). In this case, the individuals are essentially connected to others that have the same average opinions. Henceforth, we called this result as \emph{diverse} since the individuals communicate with others that have diversified opinions. A combination of the already presented density maps can form other possibilities of results. For all of the case examples, the border effect is found in the density map.

\section{Results and Discussion}
\label{sec:reults}
Here, we present the results, starting from analysis regarding the opinion distributions, by varying the post transmission parameters, reception, and rewiring. We also analyze a real scenario in terms of the proposed methodology.

\subsection{Analysis of the dynamics}
\label{sec:results_rewiring}
First, we considered the scenarios with the full dynamics, which includes the possibility of rewiring. We started by considering Erd\H{o}s-R\'enyi (ER)~\cite{erdos1960random} networks with approximately 1000 nodes and $\langle k \rangle = 8$. In this network, the connections between nodes are defined according to a probability of connection, $p$. This parameter was set to give rise to networks approximately with the desired average degree. Furthermore, the tests were also executed for $\langle k \rangle = 4$, and the results were similar to the other employed average degree (results are not shown). 

In the first tests, the network structure varies according to time, we considered a single initial structure. Additionally, we varied all possible combinations of parameters, as presented in sections~\ref{sec:post_transm_pos}~and~\ref{sec:post_distributions_pos}. In the case of phase $\phi$, we considered 33 values between 0 and $2\pi$. The number of iterations was defined to be enough to lead the dynamics to the steady-state. More specifically, for the majority of the cases, we considered that the dynamics reached steady-state when there are no significant variations of $BC$ along time. For more details, see Supplementary material~\ref{sec:temporal}.

\begin{figure*}[!htpb]
  \centering
    \includegraphics[width=1.\textwidth]{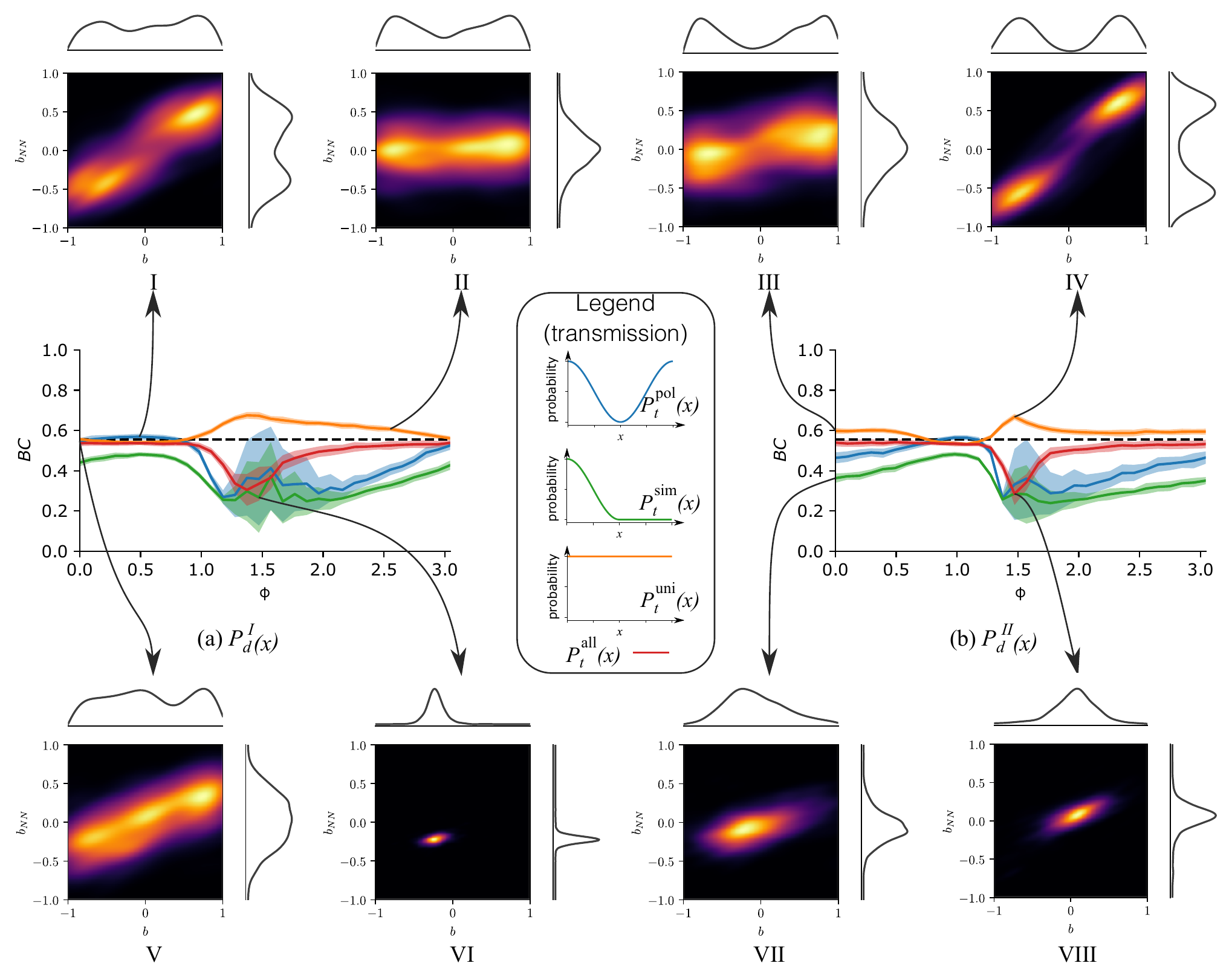}
    
  \caption{The average and respective standard deviations of $BC$ (bimodality coefficient) are shown in items (a) and (b). The horizontal dashed line indicates $BC_{\text{critc}}$. For each item, we illustrate the measurements with four examples of density maps of $b$ against $b_{NN}$. The density maps represent the following configurations: I- $P^{I}_d$, $P^{\text{pol}}_t$, and $\phi=0.49$, II- $P^{I}_d$, $P^{\text{uni}}_t$, and $\phi=2.65$, III- $P^{II}_d$, $P^{\text{uni}}_t$, and $\phi=0$, IV- $P^{II}_d$, $P^{\text{uni}}_t$, and $\phi=1.47$, V- $P^{I}_d$, $P^{\text{pol}}_t$, and $\phi=0$, VI- $P^{I}_d$, $P^{\text{sim}}_t$, and $\phi=1.47$, VII- $P^{II}_d$, $P^{\text{sim}}_t$, and $\phi=0$, and VIII- $P^{II}_d$, $P^{\text{all}}_t$, and $\phi=1.47$. These density maps illustrate typical results of a single execution of the dynamics.}
  \label{fig:complete_with_rewiring}
\end{figure*}

Many different behaviors can be observed, as illustrated in  Figure~\ref{fig:complete_with_rewiring}. For both types of distribution functions ($P^{I}_d$ and $P^{II}_d$), the type of transmission function that lead to higher values of $BC$ is $P_t^{\text{uni}}$, for values of $\phi$ close to $1.5$. This result means that, according to our model, if the social network users tend not to express a strong opinion on the posts, the algorithm (via its post distribution) can lead the opinion distributions to be polarized. By considering $P^{III}_d$, for all transmission functions, the $b$ distributions were not found to be bi-modal (more information is shown in Supplementary material~\ref{sec:equaltransmission_rewiring}). Again, the distribution function is found to play an essential role in the polarization.

Also, considering the opinion distributions, for the majority of the results, balance ($\beta$) is found to be high (more details are shown in Figure~\ref{fig:bc_x_beta_rewiring} of Supplementary material). However, for almost all combined functions, $\beta$ tends to become lower when $\phi$ is close to $\pi/2$, except for $P^{\text{uni}}_t$. For these values, the distributions can be considered unbalanced.

In this subsection, we describe the relationship between the opinion dynamics and the network topology. First, we analyze the density maps of $b$ against $b_{NN}$. We did not show average density maps because, in some cases, this average could make interesting outcomes less discernible. In general, the obtained density maps indicate results varying from consensus to echo-chambers. Items VI, VII, and VII of Figure~\ref{fig:complete_with_rewiring} illustrates three different levels of consensus, where the more well-defined scenario is found for $P^{I}_d$, $P^{\text{sim}}_t$, and $\phi=1.47$ (item VI). The echo chamber formation was found only when we employed $P_t^{\text{uni}}$, for both types of post distribution ($P_d^{I}$ and $P_d^{II}$) and $\phi$ values close to $1.5$. See an example in item IV of Figure~\ref{fig:complete_with_rewiring}. Furthermore, to lead to echo chambers, the network structure change and give rise to separated communities. However, there is no significant change in the degree distributions~\footnote{Degree is defined as the number of edges connected to a given node.}. Examples of these degree distributions are shown in Figure~\ref{fig:degree} of the Supplementary material.  

Other interesting results are shown in items I and II of Figure~\ref{fig:complete_with_rewiring}. In the first, $BC$ is slightly higher than $BC_{\text{critic}}$. So $b$'s distribution tends to be similar to a uniform distribution but with a bi-modal inclination. However, most of the samples are located in the first and third quadrants, which indicated the echo chamber's tendency. In contrast with this result, for item II of Figure~\ref{fig:complete_with_rewiring}, the opinion distribution is bi-modal, but there is no tendency of echo chambers. This density map is more similar to diverse, but with a bi-modal in $b$ distribution.

\subsection{Analysis of the dynamics without rewiring}

In order to better understand our proposed model, we test the dynamics without the possibility of echo chambers. Similarly to section~\ref{sec:results_rewiring}, here we consider many different combinations of parameters. As well as in the first case, we analyzed the steady-state of the dynamics. In this section, we summarize our main findings for the dynamics without rewiring in the last step of the process. More details are left in supplementary information~\ref{sec:without_rewiring_dynamics}. 

First, we employ the uniform version of the post distribution. In this case, the resulting opinion distributions reflect the post transmission functions. More specifically, for $P_t^{\text{pol}}$, $P_t^{\text{sim}}$, and $P_t^{\text{uni}}$, the dynamics converged to bi-modal, uni-modal, and uniform distributions of $b$, respectively. Interestingly, for $P_t^{\text{all}}$ the emerging $b$ distribution have an intermediate value of $BC$, which is slightly lower than $BC_{\text{critic}}$. We also fixed the function of post transmission as uniform ($P_t^{\text{uni}}$). By analyzing $P_d^{I}$, for $0 \leq \phi < \pi/2$, we found that the resulting $b$ distribution tends to be uni-modal, otherwise bi-modal. A similar result is found for $P_d^{II}$, but here the uni-modal distributions are found for $\pi/4 \leq \phi < 3\pi/4$.

By considering the combinations between post transmission and distribution, other interesting results have also been found. In contrast with the previous section, the $BC$ levels are much higher for some configurations with $P_t^{\text{pol}}$. Furthermore, when we slightly vary $\phi$, more abrupt changes of $BC$ are obtained, which can be explained by variations of the opinion distributions from uni-modal directly to bi-modal, or vice versa. These changes happen only for low values of $\beta$. More details regarding these analyses are shown in supplementary information~\ref{sec:without_rewiring_dynamics}.

Because of the large number of reception possibilities,  in the following of this subsection, we restrict the analysis to two fixed values $\phi$. These values were chosen in line with the variations of $BC$ and $\beta$ (for more information see Figure~\ref{fig:bc_x_beta} of Supplementary material). We consider $\phi$ equals to $\pi$ and $1.473$. By comparing both reception functions, in the case of $\phi=\pi$, the bimodality coefficient does not differ considerably, and high values of $\beta$ are found. One of the highest differences between $BC$ is found for $\phi=1.473$. In this case, for all tested parameters, except when we employed $P_t^{\text{uni}}$, the resulting $b$ distribution is found to be unbalanced. We compared $BC$ among several other network topologies. All in all, the results were found to be similar to the presented values (the complete analysis is available in Supplementary material~\ref{sec:comparison_topologies}).

\begin{figure}[!htpb]
  \centering
    \subfigure[$P_t^{pol}$ and $P_d^I$ with $\phi=1.473$]{\includegraphics[width=0.23\textwidth]{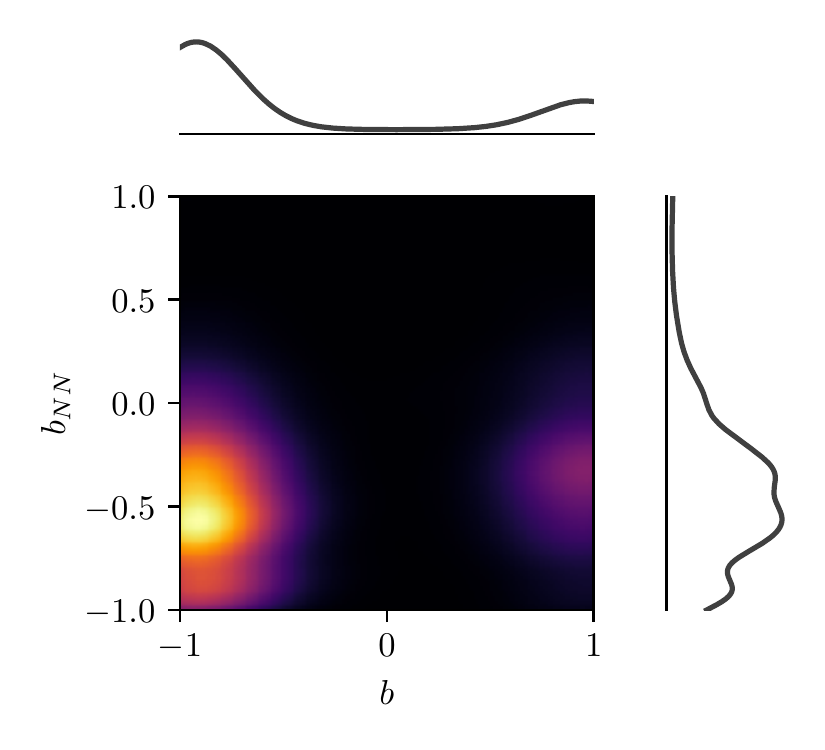}}
    \subfigure[$P_t^{pol}$ and $P_d^{II}$ with $\phi=1.473$]{\includegraphics[width=0.23\textwidth]{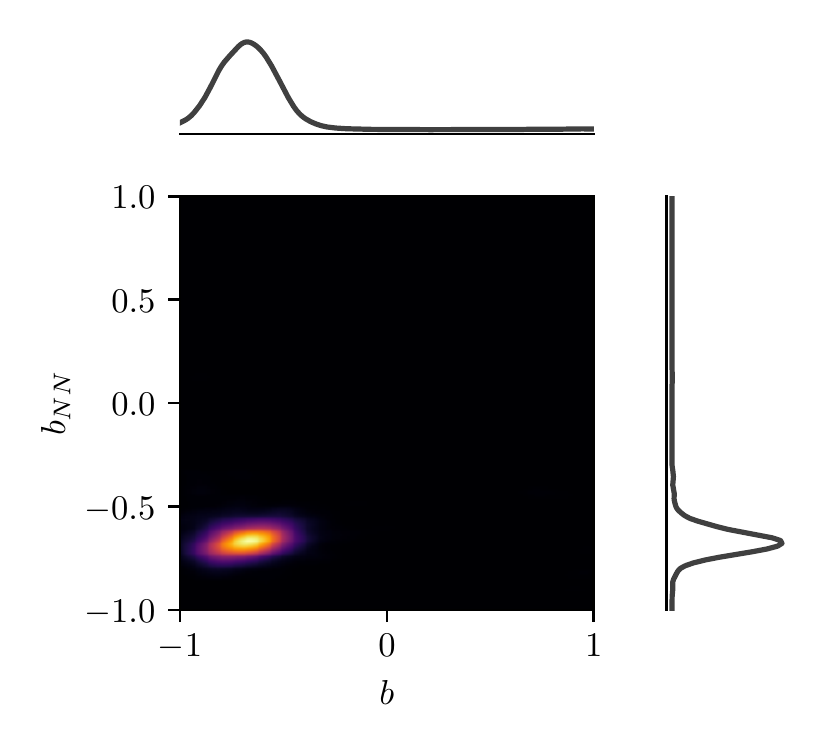}}
    \subfigure[$P_t^{sim}$ and $P_d^{II}$ with $\phi=0.0$]{\includegraphics[width=0.23\textwidth]{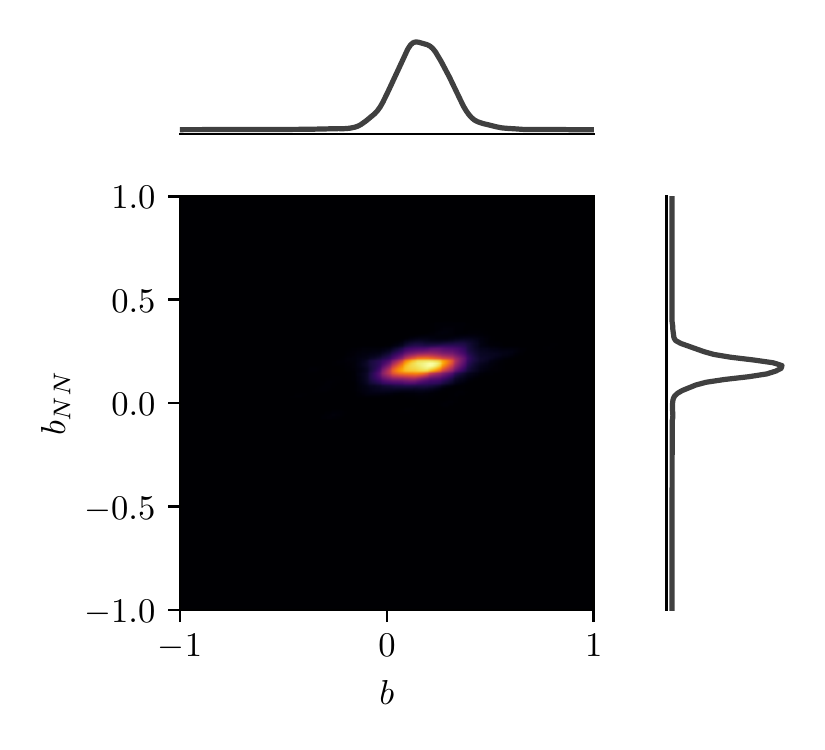}}
    \subfigure[$P_t^{pol}$ and $P_d^{II}$ with \ $\phi=0.0$]{\includegraphics[width=0.23\textwidth]{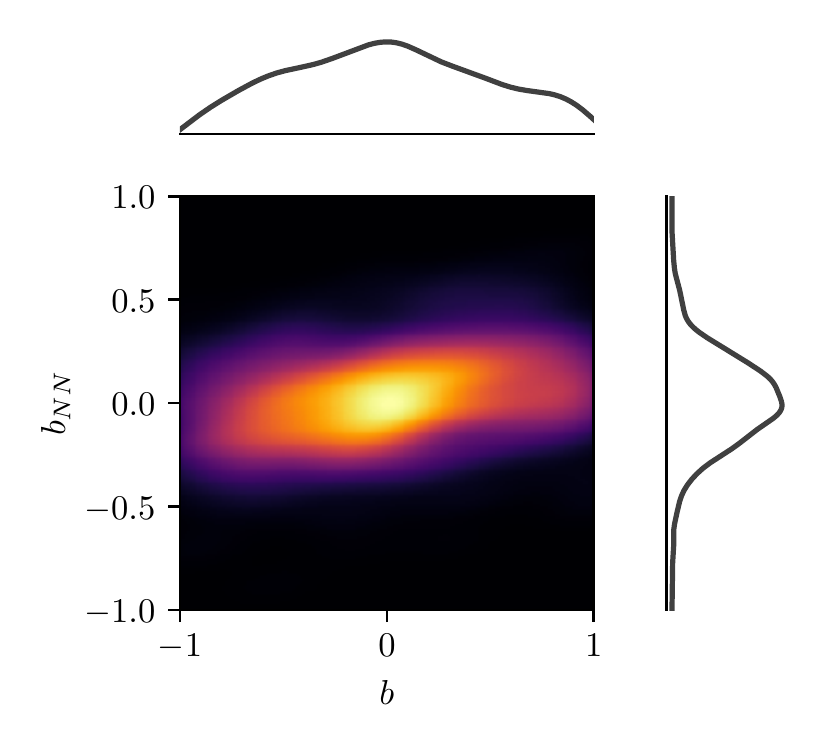}}
    
  \caption{Samples of density maps from the comparison between the $b$ against $b_{NN}$. Lighter colors represent the denser regions. All of these results were measured from ER networks with $\langle k \rangle \approx 8$, without considering the possibility of rewiring.}
  \label{fig:echoChamber}
\end{figure}

Here, we focus on the ER networks as they are the simplest and the results for other structures are similar. The different topologies, as well as the divergent results, are presented in supplementary material~\ref{sec:comparison_topologies}). Figure~\ref{fig:echoChamber} shows some examples of density maps. Figure~\ref{fig:echoChamber}(a) illustrates a scenario in which the opinions are polarized into two groups, but there are no well-defined echo chambers. More specifically, the lack of echo chamber formation is characterized by similar average opinions of the agent neighbors. In other cases, where the opinions of all of the agents are similar between themselves, a single group is found in the density map (see Figure~\ref{fig:echoChamber}(b) and Figure~\ref{fig:echoChamber}(c)). Interestingly, for the example of Figure~\ref{fig:echoChamber}(b), the opinions converge to an extreme. This result is obtained because during the transient $b$ distribution becomes bi-modal and converges to uni-modal, in which one of the two peaks increases while the other decreases, giving rise to a uni-modal distribution close to $-1$ or $1$ (this effect is shown in Figure
~\ref{fig:temporal2} of supplementary material). Furthermore, in Figure~\ref{fig:echoChamber}(d), we present an example of a density map with a tendency for the diverse scenario. However, individuals are more likely to be connected to neighbors with similar opinions. 

\begin{figure}[!htpb]
  \centering
    \subfigure[i]{\includegraphics[width=0.23\textwidth]{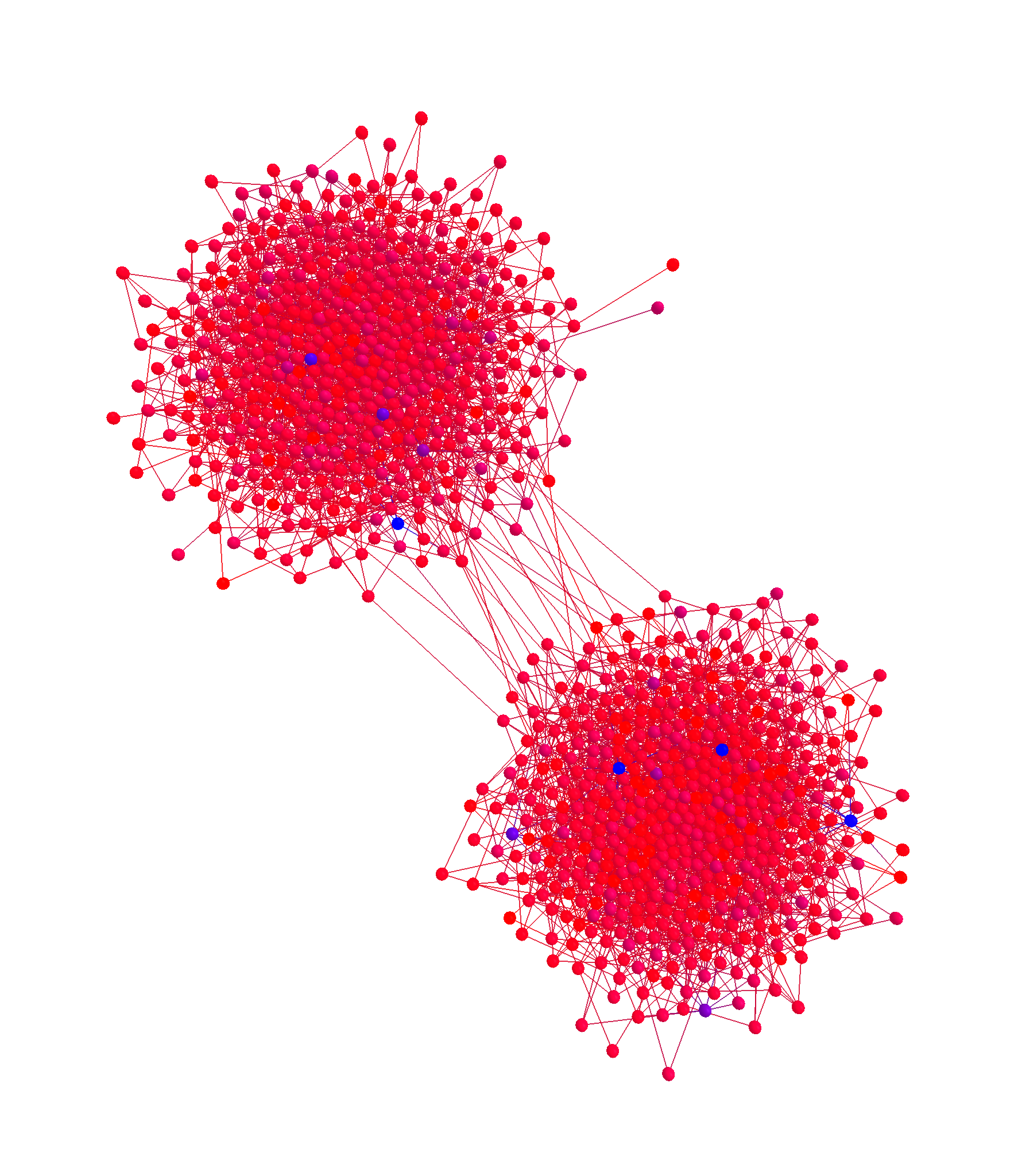}}
    \subfigure[i]{\includegraphics[width=0.23\textwidth]{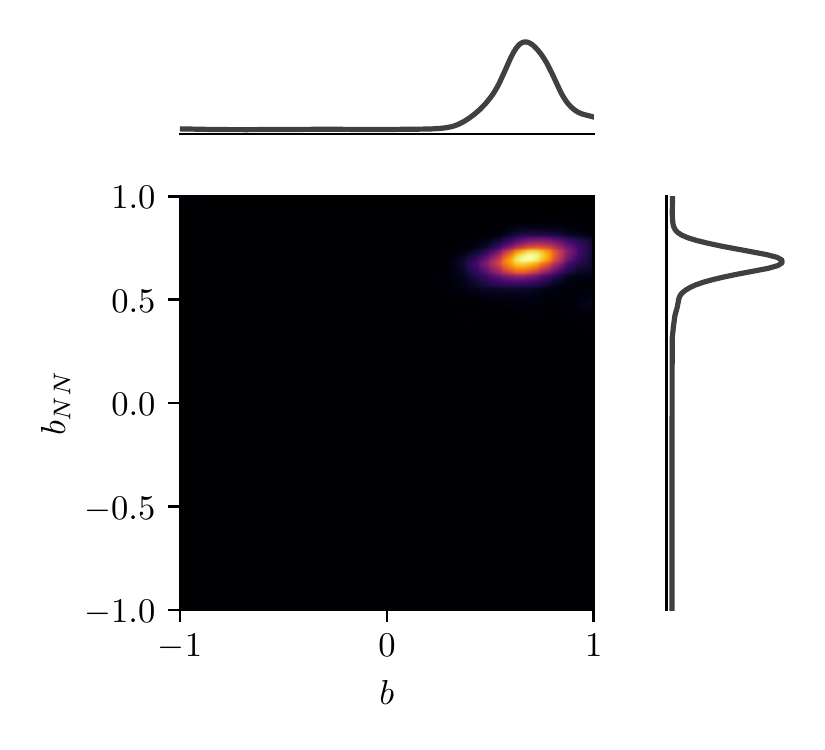}}
    \subfigure[ii]{\includegraphics[width=0.23\textwidth]{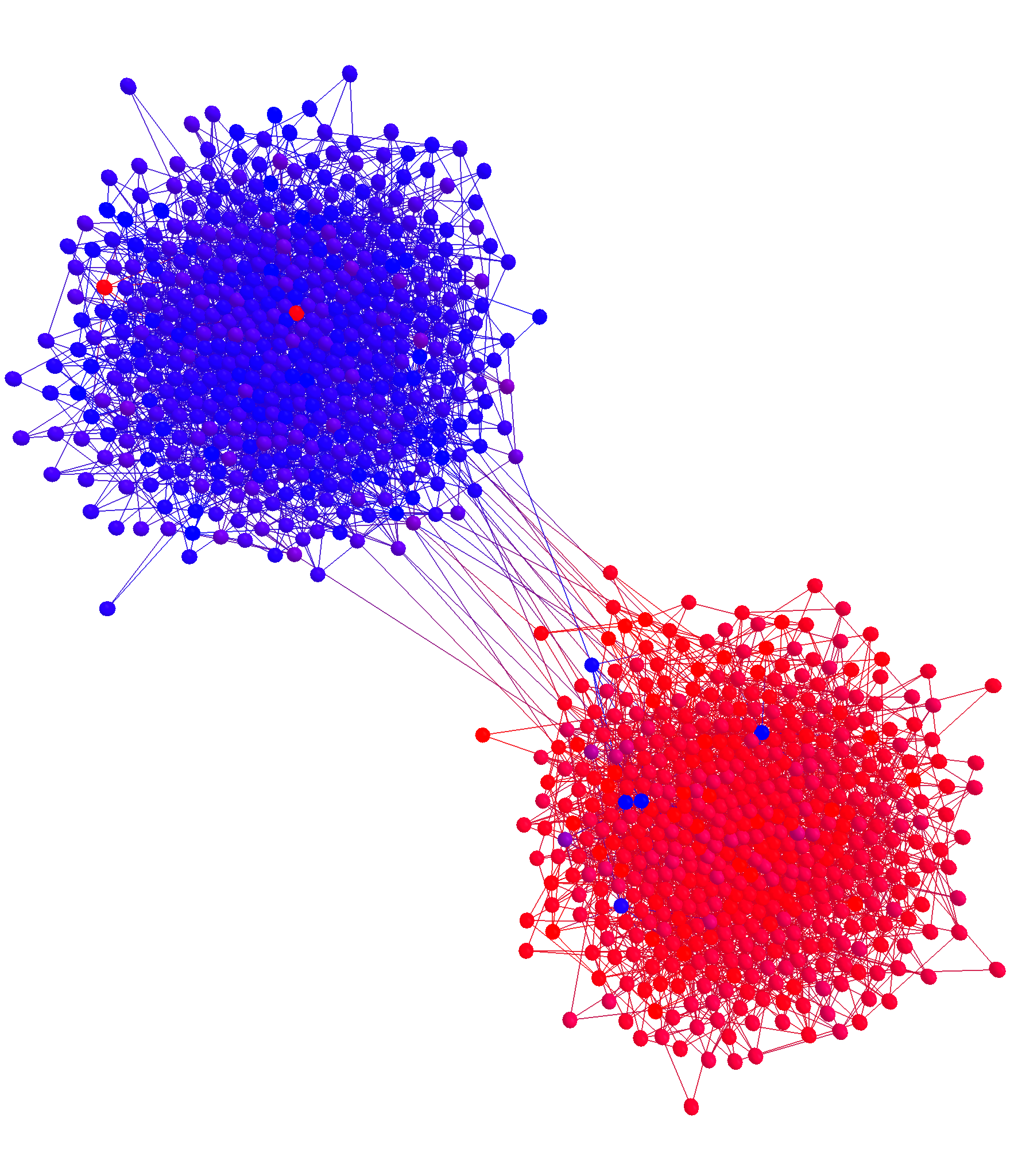}}
    \subfigure[ii ]{\includegraphics[width=0.23\textwidth]{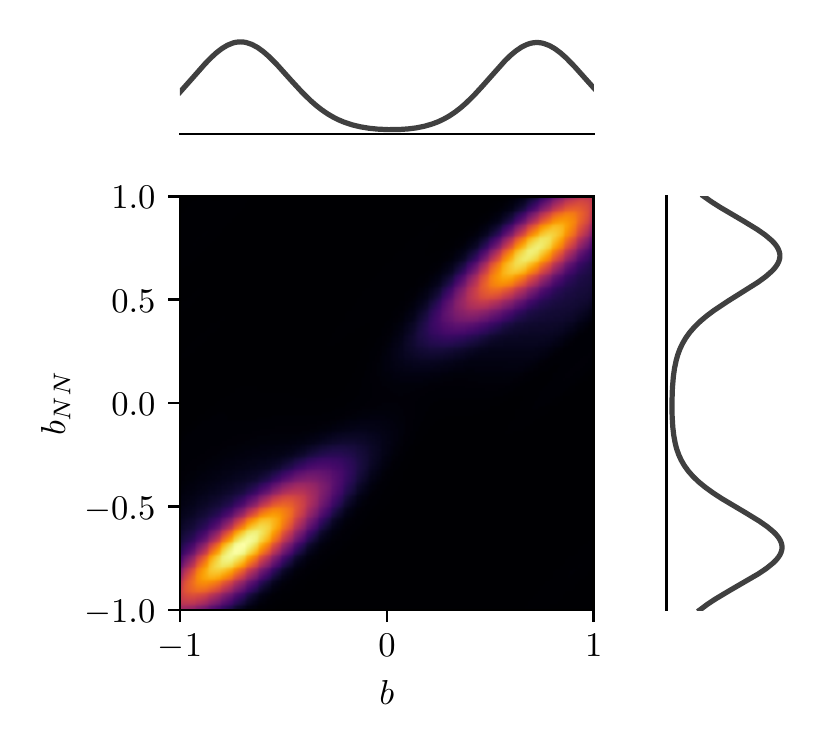}}
    
\caption{Example of two samples (i and ii) of the resulting dynamics executed on a SBM, with $P^{\text{uni}}_t$, $P^{II}_d$, and $\phi=1.47$. Items (a) and (c) display the SBM network visualizations. The colors vary from blue to red, which represent left and right wings, respectively. In (b) and (d), we display the density maps of opinions ($b$) against the average opinion of the neighbors ($b_{NN}$), where lighter colors represent large numbers.}
\label{fig:networks_sbm}
\end{figure}

Although in all results presented in this subsection, echo chambers were not found, there is the possibility to converge to echo chambers even without rewiring, depending on the network structure. In order to illustrate this possibility, we employ an SBM (Stochastic Block Model)~\cite{holland1983stochastic} network with two well-separated communities, as shown in Figure~\ref{fig:networks_sbm}. The employed probability of connection between the communities was set to $8 \times 10^{-5}$. Interestingly, this network structure leads to bistable results. In particular, the dynamics can lead to both consensuses or echo chamber formations, where $46\%$ of the 50 employed samples converged to echo chambers.

\subsection{Characterization of real data}
\begin{figure*}[!htpb]
  \centering
    \subfigure[Obamacare]{\includegraphics[width=0.3\textwidth]{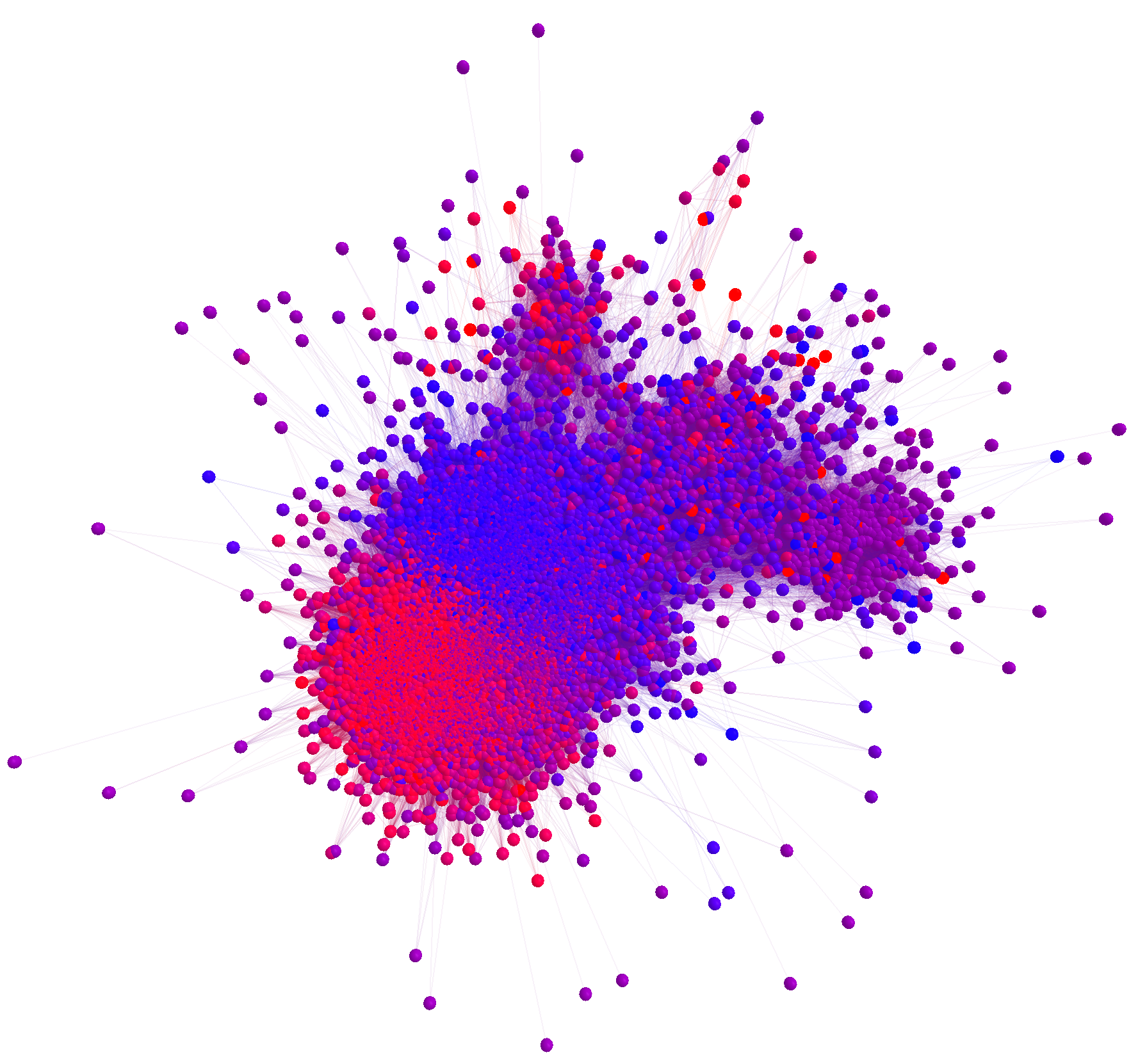}}
    \subfigure[Gun control ]{\includegraphics[width=0.3\textwidth]{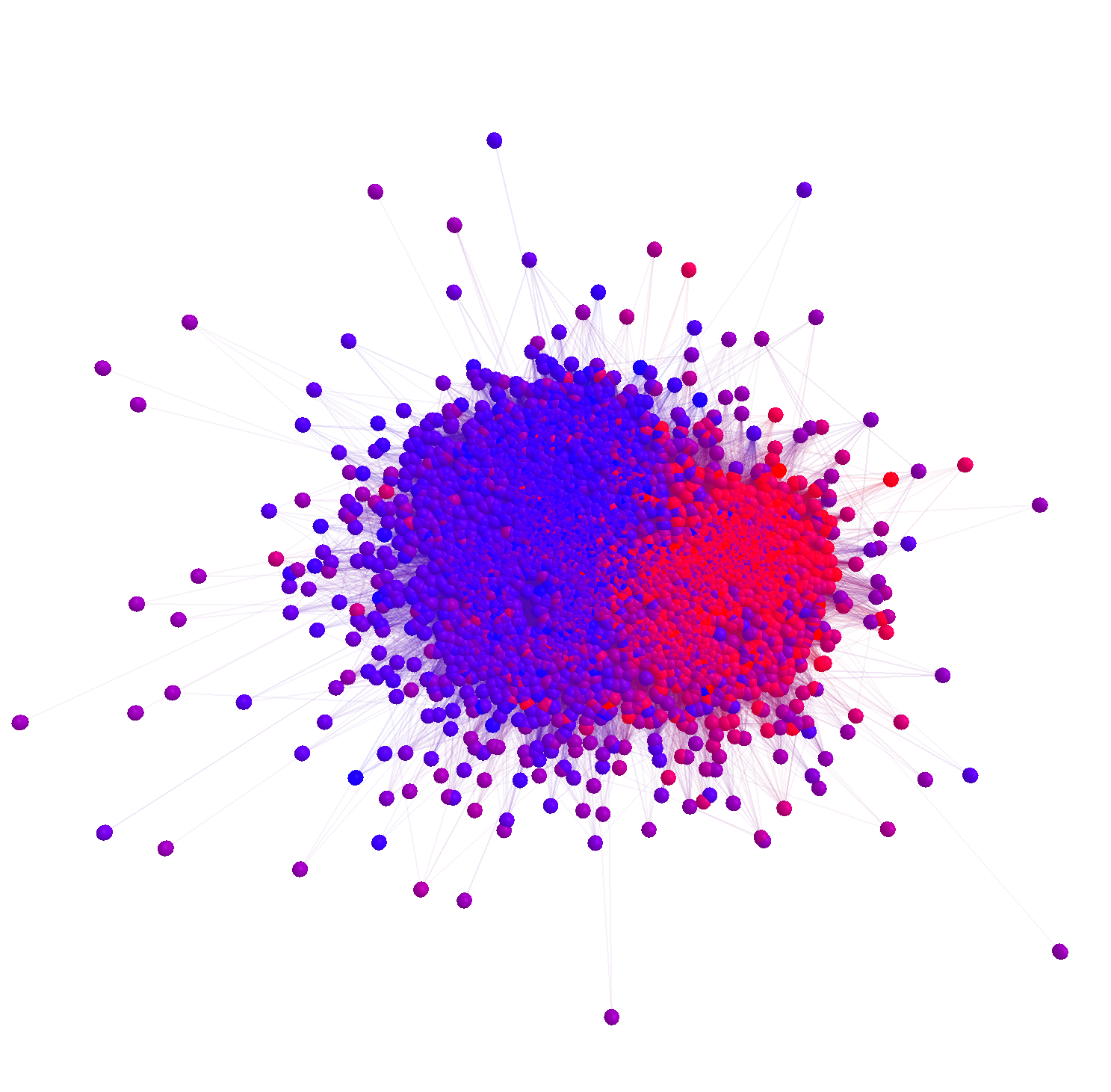}}
    \subfigure[Abortion]{\includegraphics[width=0.3\textwidth]{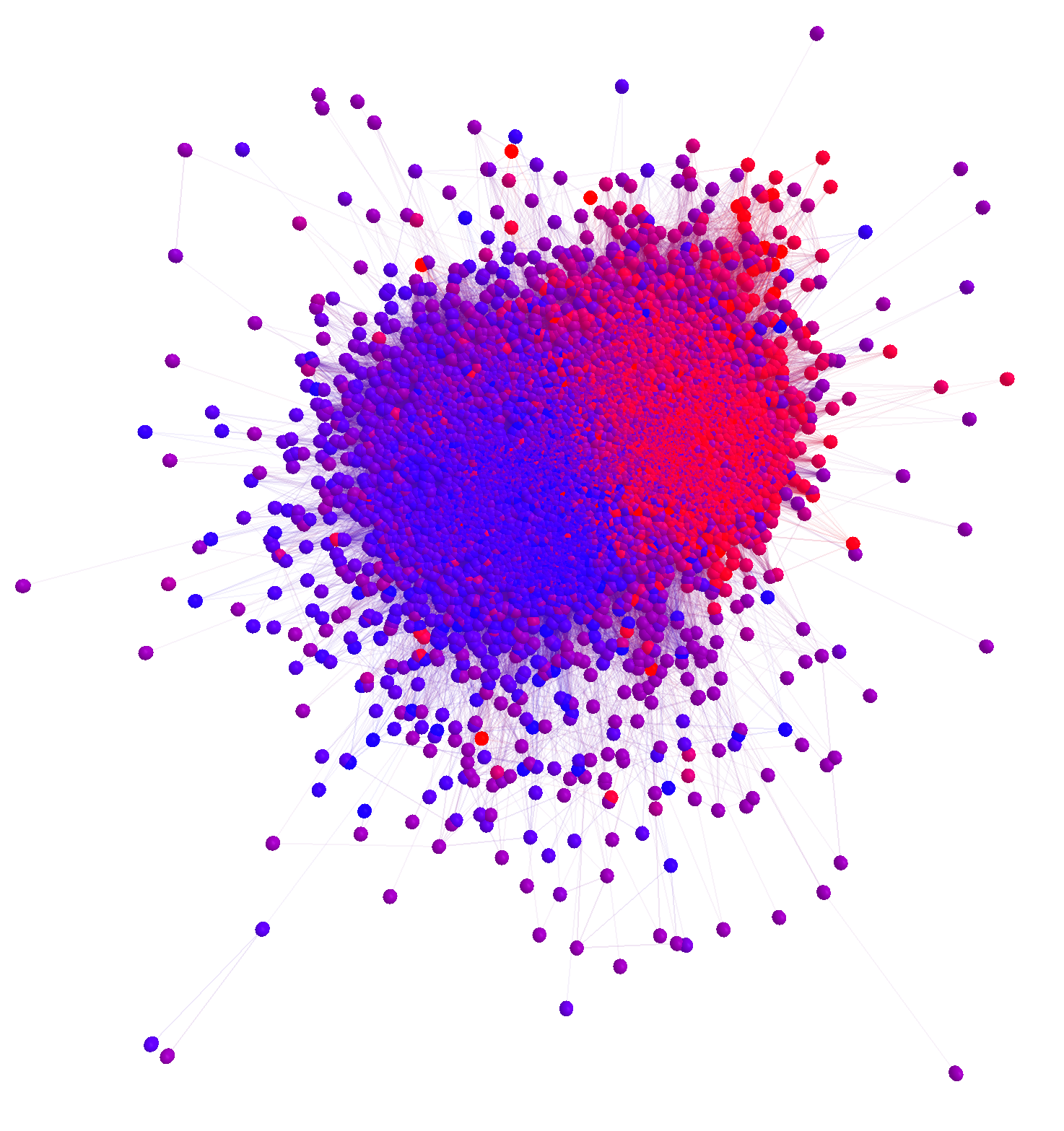}}
    
    \subfigure[Obamacare]{\includegraphics[width=0.23\textwidth]{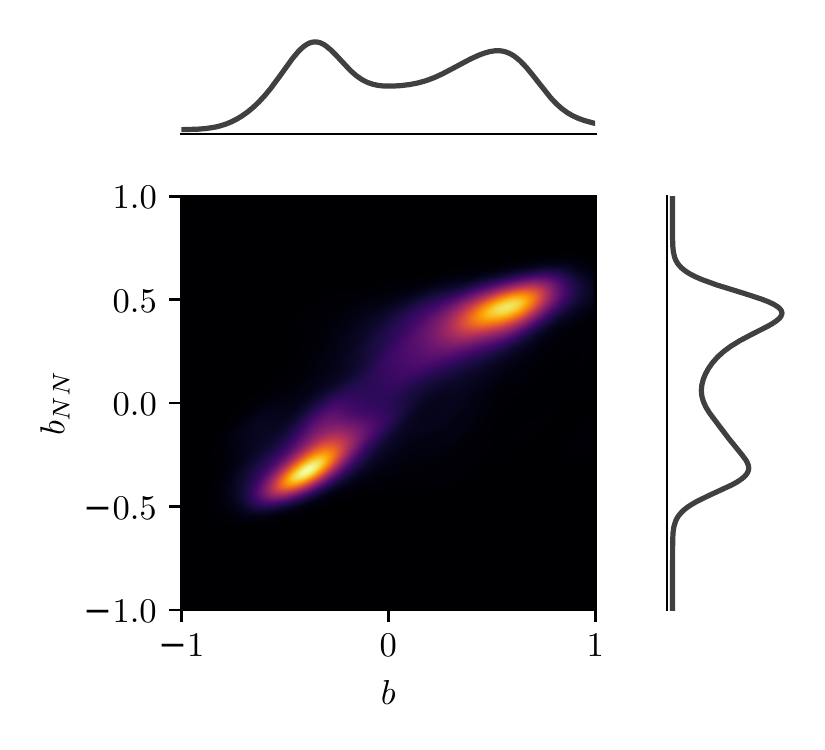}} \ 
    \subfigure[Gun control ]{\includegraphics[width=0.23\textwidth]{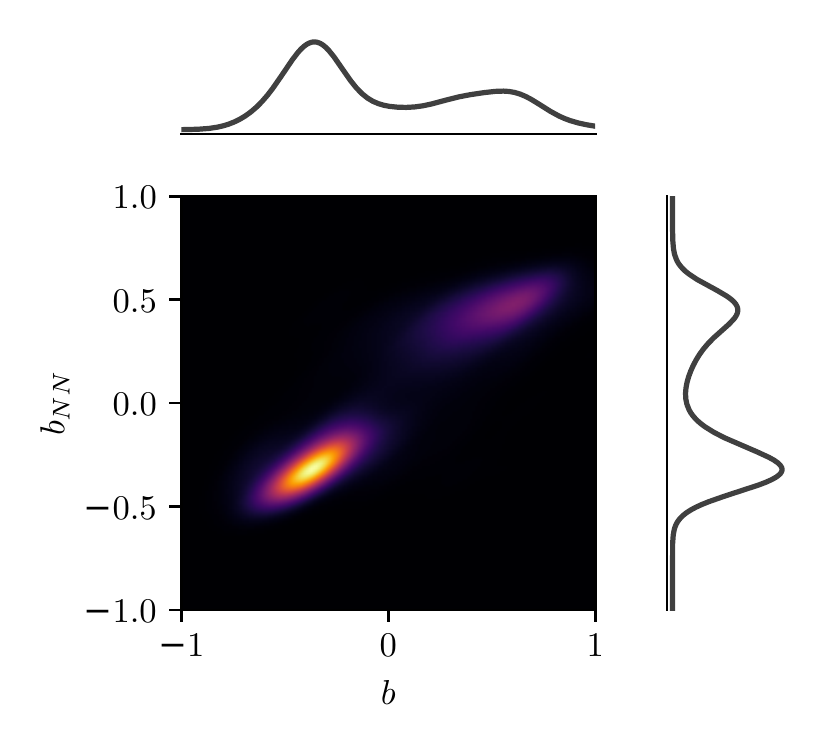}} \ 
    \subfigure[Abortion]{\includegraphics[width=0.23\textwidth]{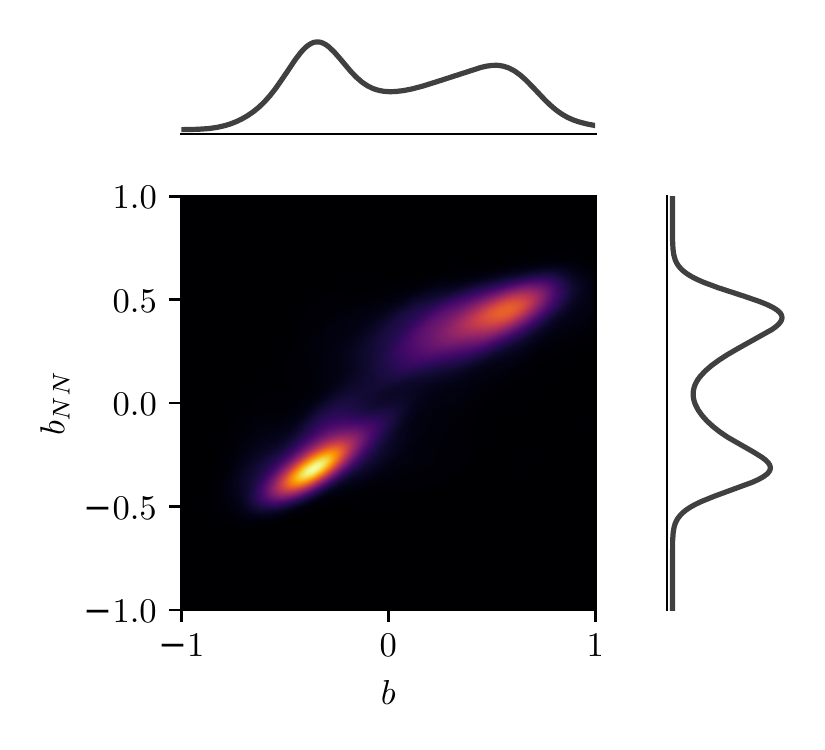}}
  \caption{Real data visualizations. The three first panels display the Twitter network visualizations, in which the colors vary from blue to red. More specifically, blue and red represent left and right wings, respectively. The second line shows the respective density maps of opinions ($b$) against the average opinion of the neighbors ($b_{NN}$), in which lighter colors represent large numbers.}
  \label{fig:networks}
\end{figure*}

We employed networks obtained from Twitter that represent political examples of polarization in the United States, obtained in~\cite{garimella2018political}, and studied in~\cite{cinelli2020echo}. The considered subjects are: Obamacare (8703 nodes and 3,797,871 edges), gun control (3963 nodes and 1,053,275 edges), and abortion (7401 nodes and 2,330,276 edges). In all networks, nodes represent users, and the directed connections were created according to followers (from following to follower). See the real data visualizations in Figure~\ref{fig:networks}. Furthermore, online news organizations with political inclinations were used to define the individuals' opinions~\cite{cinelli2020echo}. For the sake of simplicity, in our analysis, we considered the network as being undirected.

The opinions of the network users, shown in Figures~\ref{fig:networks}(a),~(b),~and~(c), seem to be separated. To compare these opinion distributions with the previous results, we reproduce the previously employed measurements. Table~\ref{tab:real_net_information} presents the measurements of $BC$ and $\beta$. Comparing the real data with our previous experiments, we observe that in the cases in which we considered rewiring, $BC$ was similar to the real networks' measures (see for instance Figure~\ref{fig:complete_with_rewiring}, panel IV). However, we remark that $\beta$ is lower for real cases than for our simulation. The abortion network is the case in which our model better approximates real data. Thus, suggesting that our model can be helpful both in a quantitative and qualitative analysis of real systems, providing additional interpretations of real phenomena.

\begin{table}[!htpb]
    \centering
    \begin{tabular}{|l|c|c|}
    \hline
    Subject & $BC$ & $\beta$ \\ \hline \hline
    Obamacare & 0.60 & 0.80\\
    Gun control & 0.67 & 0.70\\
    Abortion & 0.60 & 0.91\\
    \hline
    \end{tabular}
    \caption{Measures of bimodality coefficient, $BC$, and balance, $\beta$, obtained from real Twitter networks.}
    \label{tab:real_net_information}
\end{table}

\section{Conclusions}
\label{sec:concusions}
Due to the rise of social network users, researchers have been studying respective opinion dynamics. Here, we proposed a model to study the configurations that can give rise to polarization. Our dynamics is based on some compartmentalized modules, and it is limited to simulating how new information is generated, as well as the individual's friends' reactions. First, in the post transmission, the user has contact with an external piece of information and chooses if he/she will post it according to a given probability. In the post distribution, the piece of information is analyzed by the social network algorithm. The opinions of the individuals that receive the post can be attracted or repulsed by its content. For the repulsed individuals, a probability function controls if the individual will rewire friendship.  Furthermore, this study does not contribute only to a proposed model but also proposes a new type of analysis. Here, we considered that the network topology is not the single aspect that limits the communication between individuals. 

Several interesting outcomes have been observed. For instance, when we considered the function of rewiring probability, only with the uniform transmission, the dynamics gave rise to echo chambers. According to our model, if the users do not mind about the information they post, the polarization and formation of echo chambers can be influenced by the distribution function. Furthermore, three opinion organizations have been observed: consensus, echo chamber, and diverse.   

In some cases, high values of balance and low values of the bimodality coefficient have been found. This result means that the dynamics converged to consensus, but with average values close to $-1$ or $+1$, which mainly happened when we considered the dynamics without the possibility of rewiring. Also, without including rewirings, polarization can be observed for a wider range of configurations. However, for the majority of these polarized cases, there is no echo chamber formation. One exception is a network with well-separated communities, which can converge to both consensuses or echo chambers for the same set of parameters.

The bimodality coefficient was illustrated with respect to synthetic and real data, with similar balance values being obtained, especially in the case of the Abortion data.

One of our model's current limitations is that the individuals take a look at all received posts. In future works, a probability could be associated with this action, allowing posts to be discarded. Post distribution could also be adaptive and change over time.  Furthermore, we considered only undirected networks, but our model can also be implemented using directed structures.

\section*{Acknowledgments}
The authors thanks Michele Starnini for sharing the Twitter networks.
Henrique F. de Arruda acknowledges FAPESP for sponsorship (grants 2018/10489-0 and 2019/16223-5). 
Guilherme F. de Arruda and Yamir Moreno acknowledge support from Intesa Sanpaolo Innovation Center. Luciano da F. Costa thanks CNPq (grant no. 307085/2018-0) and NAP-PRP-USP for sponsorship. 
Yamir Moreno acknowledges partial support from the Government of Arag\'on and FEDER funds, Spain through grant ER36-20R to FENOL, and by MINECO and FEDER funds (grant FIS2017-87519-P). 
Research carried out using the computational resources of the Center for Mathematical Sciences Applied to Industry (CeMEAI) funded by FAPESP (grant 2013/07375-0). This work has been supported also by FAPESP grants 2015/22308-2.
The founders had no role in study design, data collection, and analysis, decision to publish, or preparation of the manuscript.

\bibliography{refs}

\section*{Supplementary material}
\renewcommand{\thefigure}{S\arabic{figure}}
\setcounter{figure}{0}

\renewcommand{\thesection}{S\arabic{section}}
\setcounter{section}{0}

\section{Transient analysis}\label{sec:temporal}
\begin{figure}[!htpb]
  \centering
    \subfigure[$P^I_d$]{\includegraphics[width=0.42\textwidth]{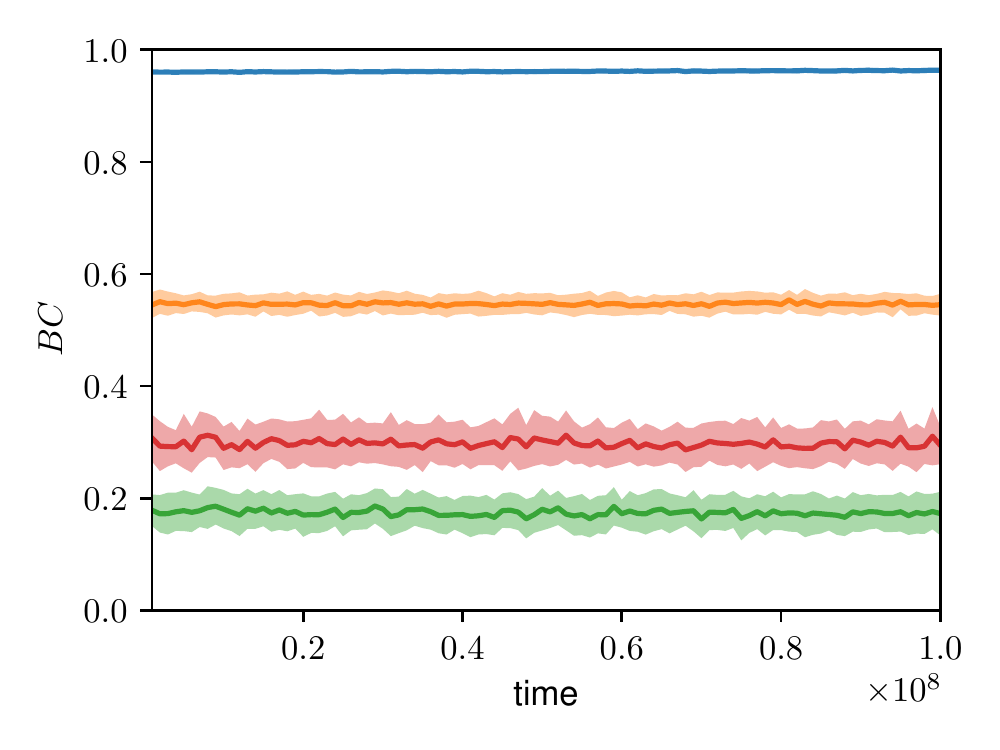}}
    \subfigure[$P^{II}_d$]{\includegraphics[width=0.42\textwidth]{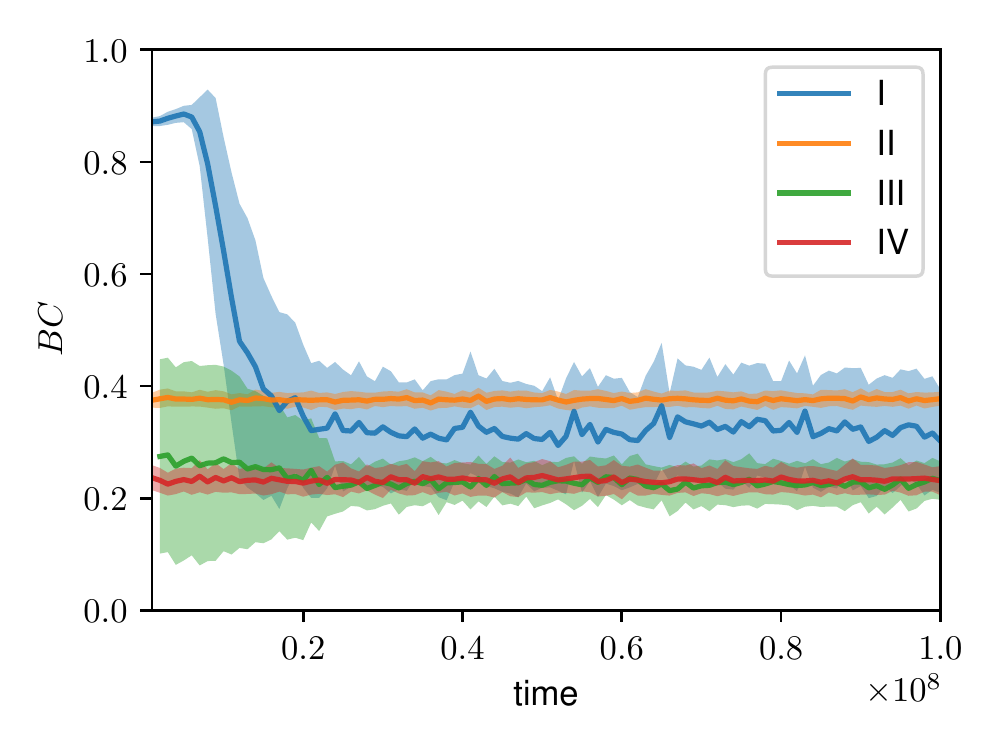}}

  \caption{Variation of BC according to the dynamics execution. The colors represent transmission probability functions, as follows: I- $P_t^{\text{pol}}(x)$ (blue), II- $P_t^{\text{uni}}(x)$ (orange), III- $P_t^{\text{sim}}(x)$ (green), and IV- $P_t^{\text{all}}(x)$ (red). The shaded regions account for standard deviations. Note that these curves start in one million iterations.}
  \label{fig:temporal}
\end{figure}

\begin{figure}[!htpb]
  \centering
    \includegraphics[width=0.5\textwidth]{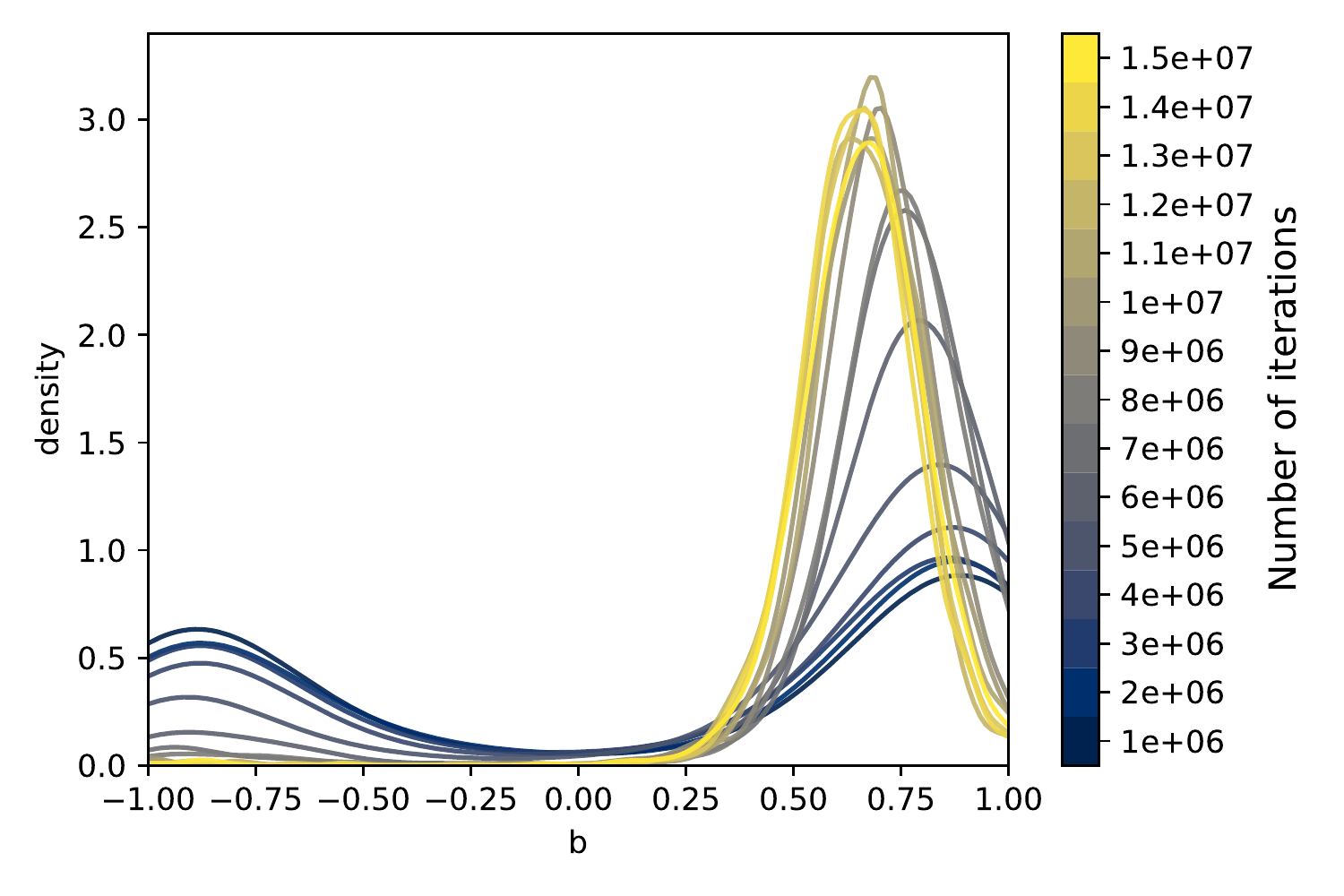}

  \caption{Temporal evolution of the opinion distribution for one execution of the dynamics, by considering the following parameters: $P_t^{pol}$ and $P_d^{II}$ with $\phi=1.473$. The first curve represents the execution with one million iterations.}
  \label{fig:temporal2}
\end{figure}

We study the evolution of $BC$ (bimodality coefficient) according to time. Figure~\ref{fig:temporal} illustrates the case of the ER network for $\langle k \rangle \approx 8$, in which the possibility of rewiring is not considered. For most of the tests (considering or not rewiring), the results were similar. Interestingly, for $P_t^{\text{pol}}$ and $P_d^{II}$ with $\phi=1.473$, BC increases to values close to 0.9 becoming highly bi-modal, and decreases to converges as being uni-modal (see Figure~\ref{fig:temporal}(b)). The evolution of one execution of its dynamics is shown in Figure~\ref{fig:temporal2}, which shows that after the first million iterations, the opinion's distribution tends to be organized as bi-modal. In the following iterations, this bi-modal distribution changes to become uni-modal, where the individual's opinions migrate from one extreme to the other. In the example shown in Figure~\ref{fig:temporal2}, individuals with negative opinions change to positive. 

\begin{figure}[!htpb]
  \centering
    \includegraphics[width=0.5\textwidth]{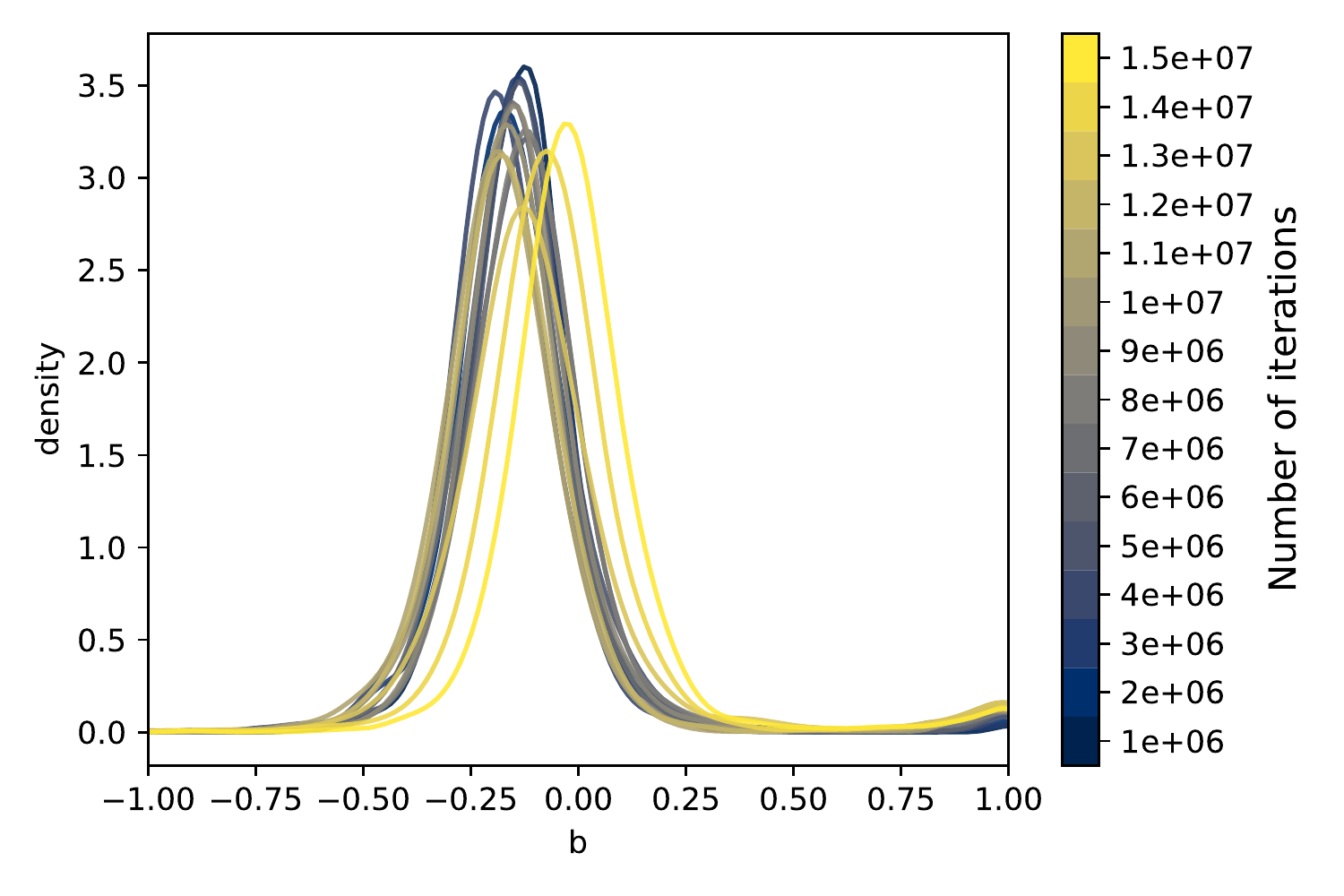}

  \caption{Temporal evolution of the opinion distribution for a single execution including the rewiring function ($P_t^{\text{pol}}$ and $P_d^{I}$ with $\phi=1.473$). The first curve represents the execution with one million iterations. The peak close to 1 can lead $BC$ to indicate that the distribution is uni-modal or bi-modal, depending on its intensity.}
  \label{fig:temporal_rewire}
\end{figure}

When we considered the possibility of rewiring (equation~\ref{eq:rewiring}), for some sets of parameter combinations, $BC$ cannot be used to account for the dynamics convergence. In these cases, we visualize the temporal evolution of different executions' steps (see an example in Figure~\ref{fig:temporal_rewire}). Interestingly, $BC$ changes abruptly because the opinion distribution goes directly from uni-modal to bi-modal without becoming uniform. This effect happens because $BC$ only quantifies the shape of the distribution (between bi-modal and uni-modal) but does not account for the peak sizes.

\section{Analysis of bimodality}
\label{sec:analysis_of_bimodality}
In order to better understand how each of the model steps is influencing the opinion distributions, we include more information regarding the dynamics executions. We also executed the dynamics by removing some parts. So, we could understand the influence of each of the parts in the dynamics executions. 

\subsection{Rewiring dynamics}
\label{sec:equaltransmission_rewiring}
\begin{figure}[!htpb]
    \centering
     \includegraphics[width=.45\textwidth]{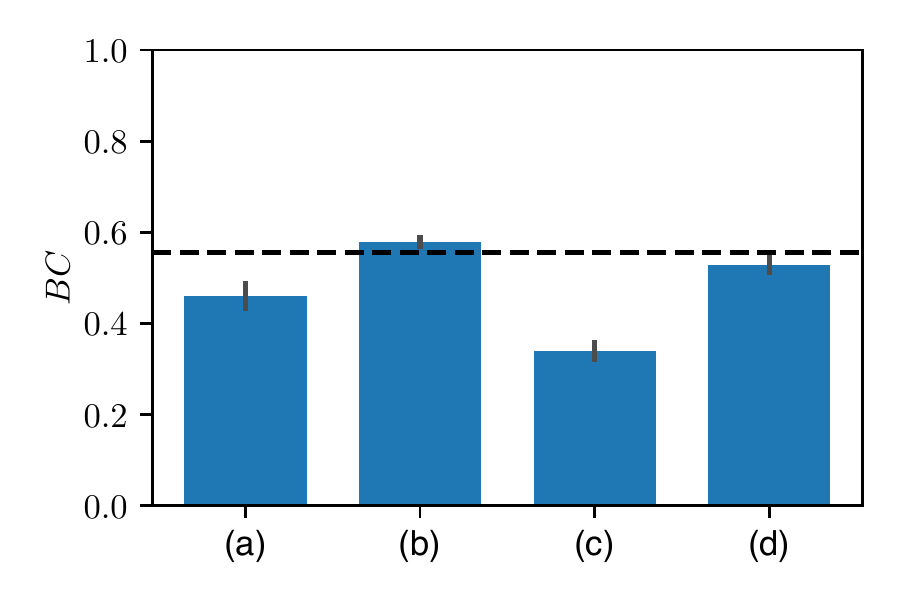}
   \caption{Average and standard deviations of the resulting $b$ distribution measured from a fixed reception probability, $P_d^{III}$ (null model) and with the rewiring function. The employed transmission probabilities are listed as follows: (a) $P_t^{\text{pol}}(x)$, (b) $P_t^{\text{uni}}(x)$, (c) $P_t^{\text{sim}}(x)$, and (d) $P_t^{\text{all}}(x)$.}
  \label{fig:equaltransmission_rewiring}
\end{figure}

\begin{figure}[!htpb]
  \centering
    \subfigure[$P_t^{pol}(x)$]{\includegraphics[width=0.23\textwidth]{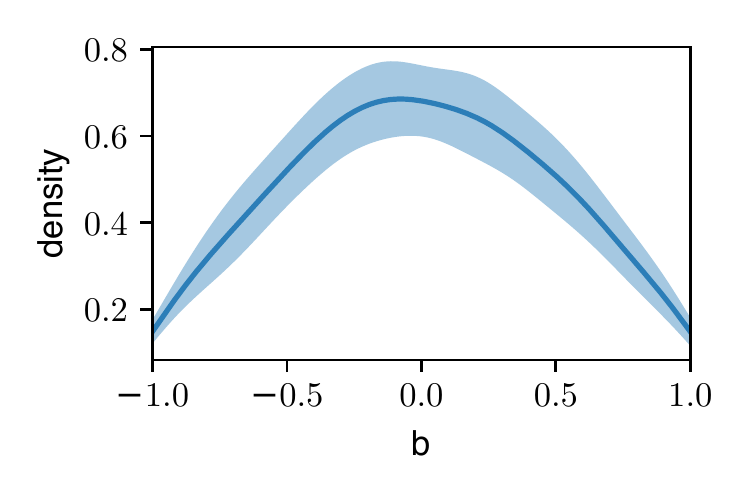}}
    \subfigure[$P_t^{uni}(x)$]{\includegraphics[width=0.23\textwidth]{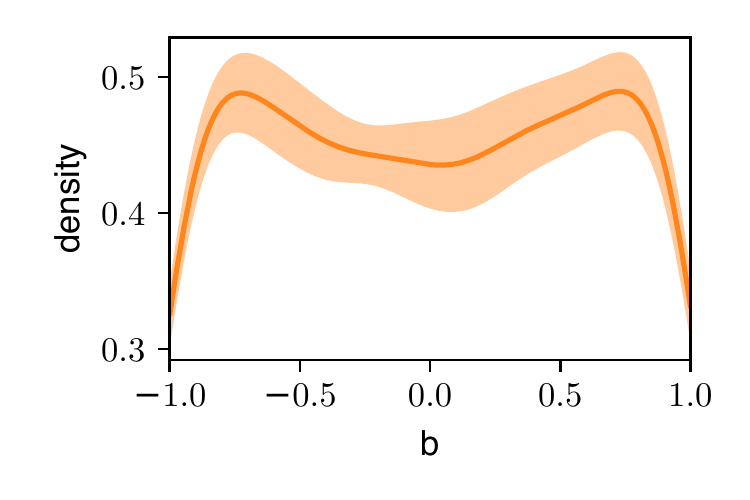}}
    \subfigure[$P_t^{sim}(x)$]{\includegraphics[width=0.23\textwidth]{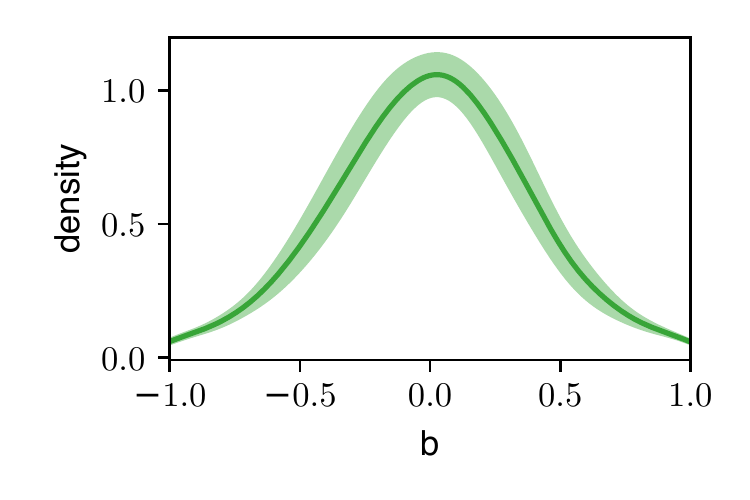}}
    \subfigure[$P_t^{all}(x)$]{\includegraphics[width=0.23\textwidth]{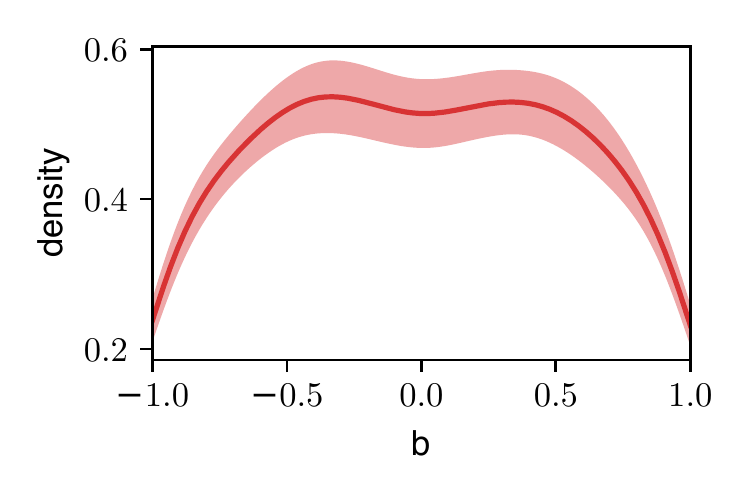}}
    
  \caption{Average opinion distributions obtained from the execution of the dynamics with a fixed reception probability ($P_d^{III}$). Furthermore, these executions considered the rewiring function. Each item represents a different transmission function. The shaded regions show the standard deviations.}
  \label{fig:avgOpEqTransm_rewiring}
\end{figure}

\begin{figure*}[!htpb]
  \centering
     \subfigure[$P^{I}_d$ and $P_t^{pol}(x)$ ]{\includegraphics[width=0.24\textwidth]{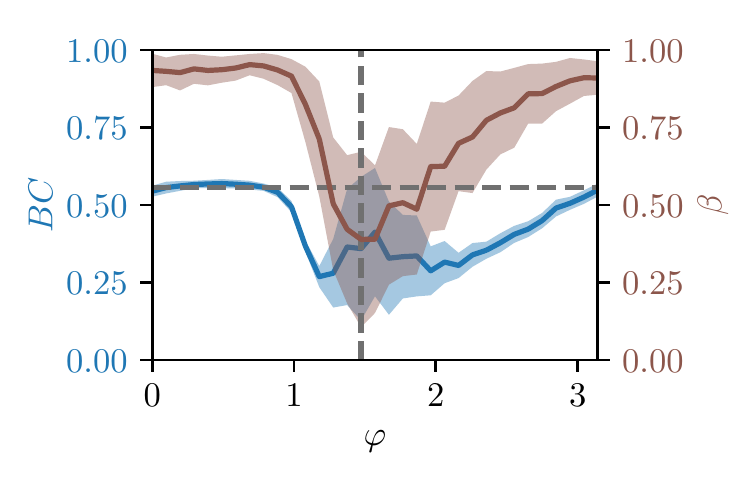}}
    \subfigure[$P^{I}_d$ and $P_t^{uni}(x)$ ]{\includegraphics[width=0.24\textwidth]{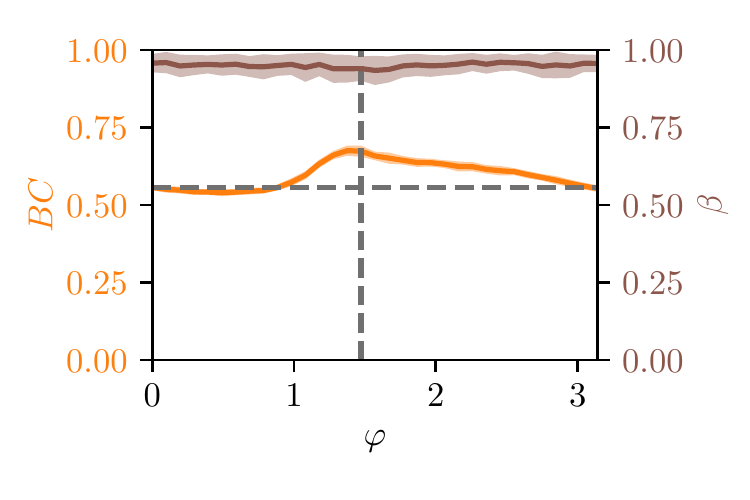}}
    \subfigure[$P^{I}_d$ and $P_t^{sim}(x)$ ]{\includegraphics[width=0.24\textwidth]{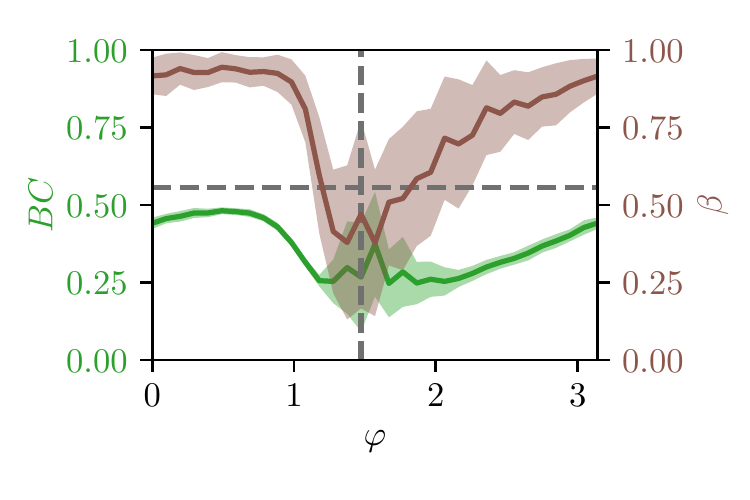}}
    \subfigure[$P^{I}_d$ and $P_t^{all}(x)$ ]{\includegraphics[width=0.24\textwidth]{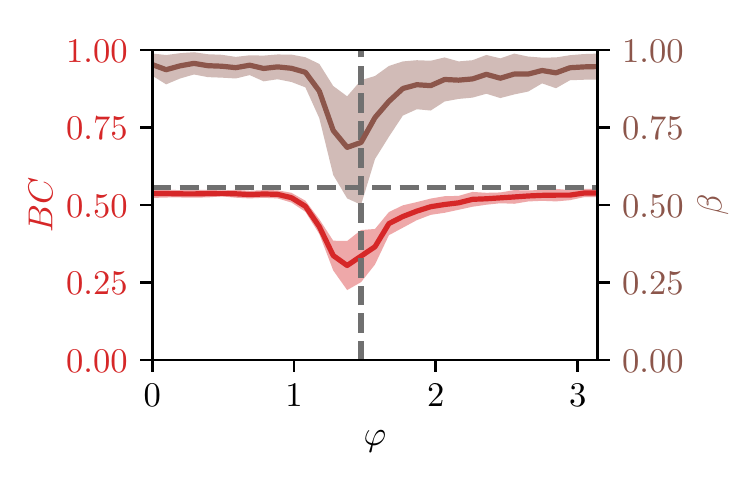}}
    
    \subfigure[$P^{II}_d$ and $P_t^{pol}(x)$ ]{\includegraphics[width=0.24\textwidth]{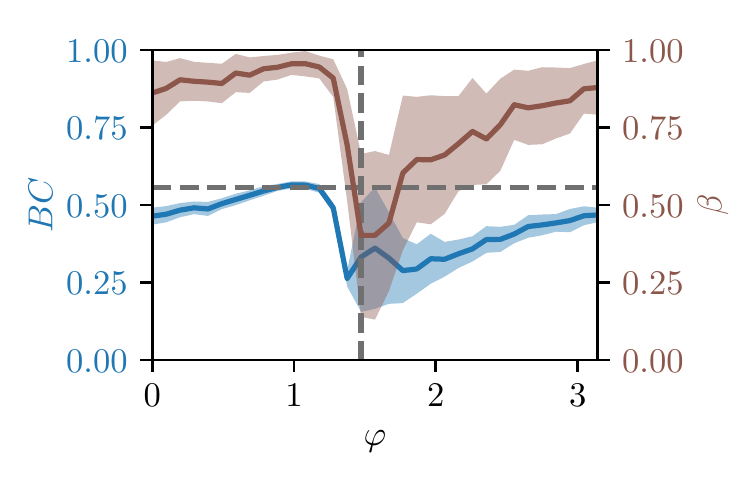}}
    \subfigure[$P^{II}_d$ and $P_t^{uni}(x)$ ]{\includegraphics[width=0.24\textwidth]{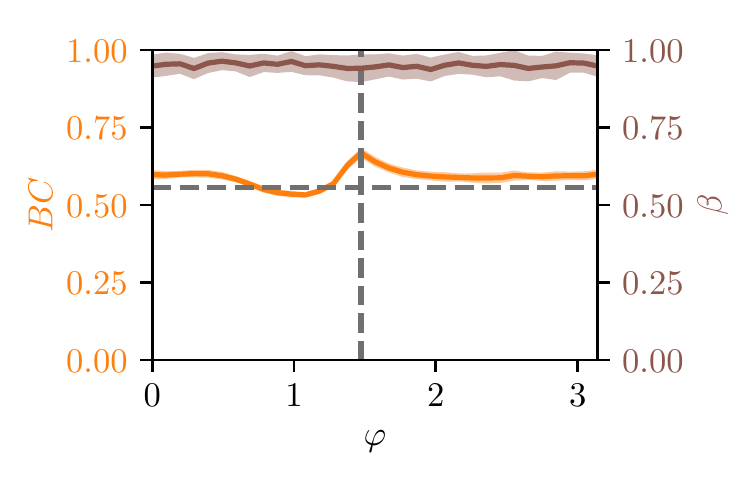}}
    \subfigure[$P^{II}_d$ and $P_t^{sim}(x)$ ]{\includegraphics[width=0.24\textwidth]{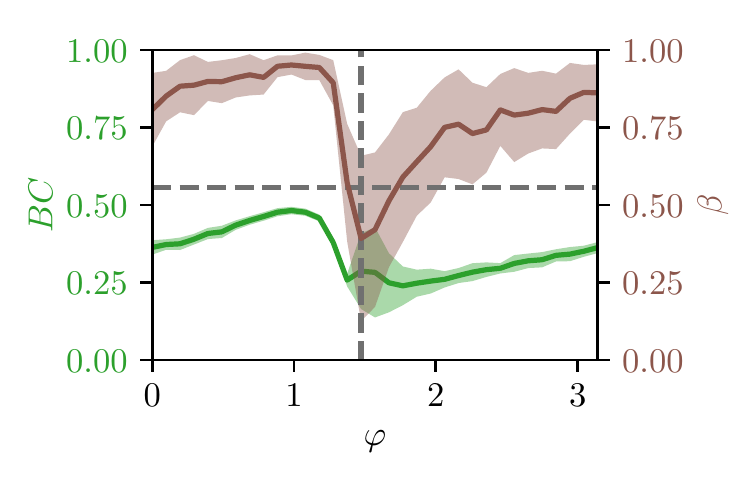}}
    \subfigure[$P^{II}_d$ and $P_t^{all}(x)$ ]{\includegraphics[width=0.24\textwidth]{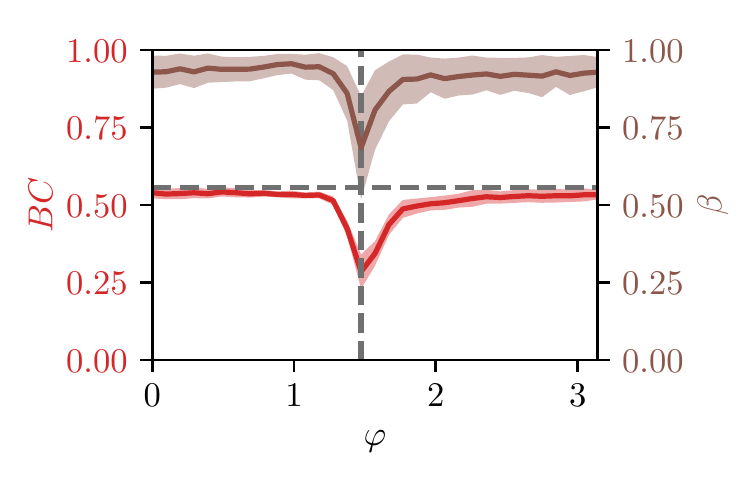}}
    
  \caption{Comparison between $BC$ and $\beta$ (in brown), for all tested parameters, and considering the possibility of rewiring. The shaded area represents the standard deviations.}
  \label{fig:bc_x_beta_rewiring}
\end{figure*}

Here, we studied the separated parts of the dynamics when executed with the rewiring probability. In order to analyze the post transmission, we fixed the post distribution as uniform, $P_d^{III}$. As a result, for $P_t^{\text{pol}}$ and $P_t^{\text{sim}}$ the resulting distributions of $b$ tends to be uni-modal. In the case of and $P_t^{\text{uni}}$ and $P_t^{\text{all}}$, the dynamics resulted in uniform distributions (see Figures~\ref{fig:equaltransmission_rewiring}~and~\ref{fig:avgOpEqTransm_rewiring}). Although, for $P_t^{\text{uni}}$, $BC$ is low, in Figure~\ref{fig:avgOpEqTransm_rewiring} a weak tendency of bimodality is can be seen. This result is found probably because of the mechanism of rewiring that can happen only after a repulsion. So, individuals with divergent opinions tend to become less connected. As a complement, individuals with similar opinions tend to be connected and be attracted by their neighbors. In the following, we present the comparison between $BC$ and $\beta$, as shown in Figure~\ref{fig:bc_x_beta_rewiring}.

Figure~\ref{fig:bc_x_beta_rewiring} illustrates the comparison between $BC$ and $\beta$ for the dynamics executed with all parts and all employed combinations of parameters.

\begin{figure}[!htpb]
  \centering
    \subfigure[$P^{I}_d$, $P^{\text{uni}}_t$, and $\phi=1.47$]{\includegraphics[width=0.4\textwidth]{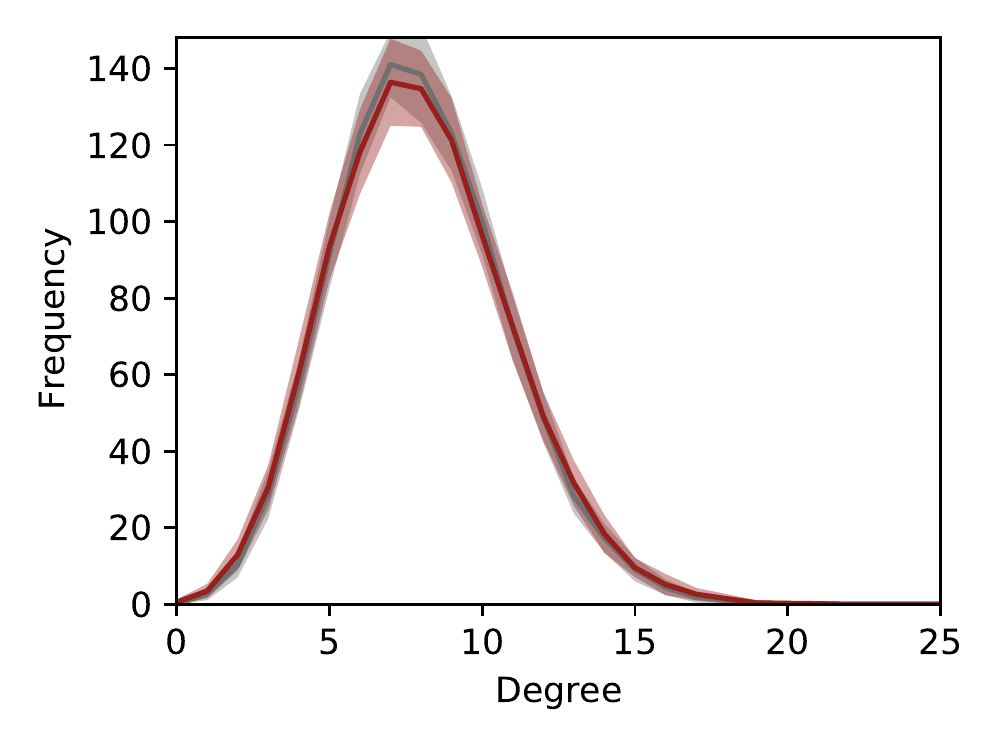}}
    \subfigure[$P^{II}_d$, $P^{\text{uni}}_t$, and $\phi=1.47$]{\includegraphics[width=0.4\textwidth]{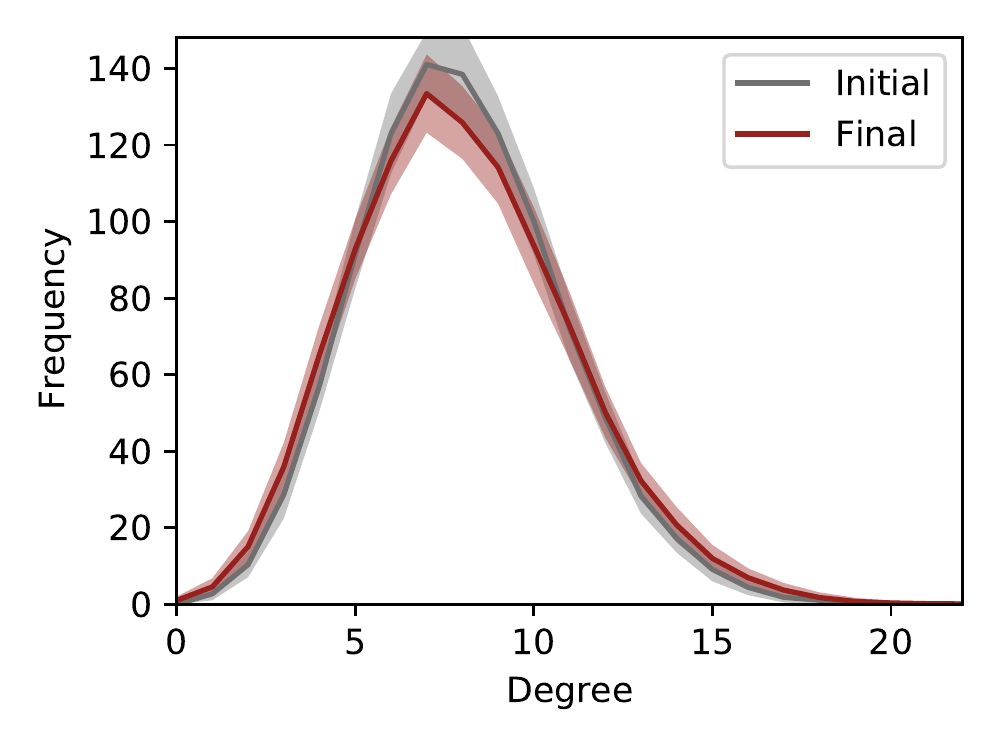}}
    
  \caption{Average degree distributions before and after the execution of the dynamics. In this case, rewiring was considered. The shaded region represents the standard deviation.}
  \label{fig:degree}
\end{figure}

In order to better understand how the topology changed according to the dynamics, we compare the degree distribution of the original with the resulting networks. This analysis was done for the case in which we observed well-defined echo chambers. Figure~\ref{fig:degree}(a) shows this comparison for $P^{I}_d$, $P^{\text{uni}}_t$, and $\phi=1.47$ and Figure~\ref{fig:degree}(b) for $P^{II}_d$, $P^{\text{uni}}_t$, and $\phi=1.47$. However, for other scenarios, the differences can vary.

\subsection{Without rewiring dynamics}
\label{sec:without_rewiring_dynamics}

\begin{figure}[!htpb]
    \centering
     \includegraphics[width=.4\textwidth]{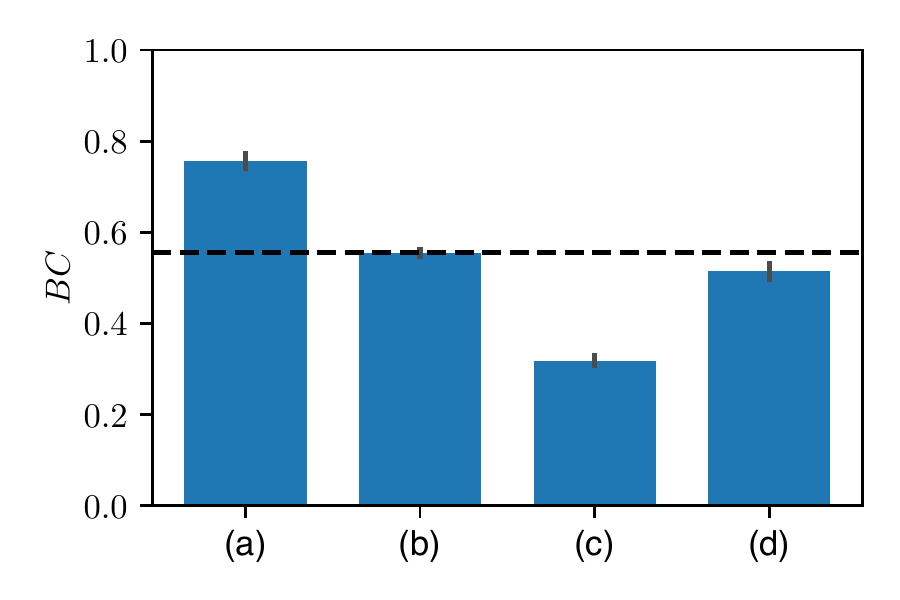}
   \caption{Average and standard deviations of the resulting $b$ distribution measured from a fixed reception probability, $P_d^{III}$. The employed transmission probabilities are listed as follows: (a) $P_t^{pol}(x)$, (b) $P_t^{uni}(x)$, (c) $P_t^{sim}(x)$, and (d) all functions with the same probability.}
  \label{fig:equaltransmission}
\end{figure}

\begin{figure*}[!htpb]
  \centering
    \subfigure[$P_t^{pol}(x)$]{\includegraphics[width=0.24\textwidth]{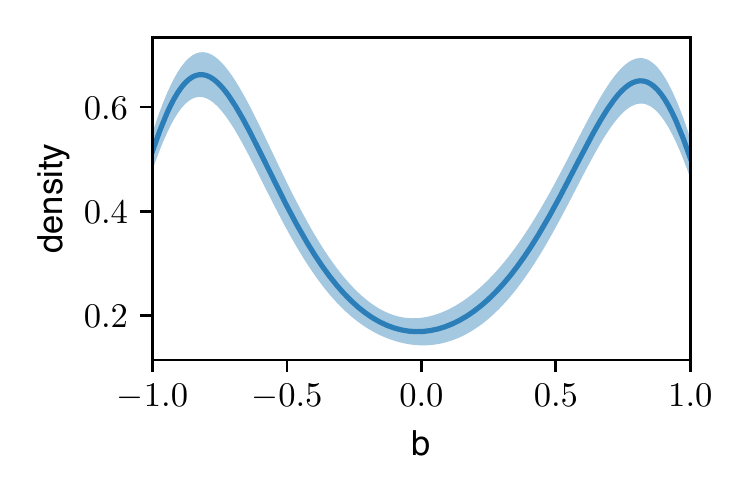}}
    \subfigure[$P_t^{uni}(x)$]{\includegraphics[width=0.24\textwidth]{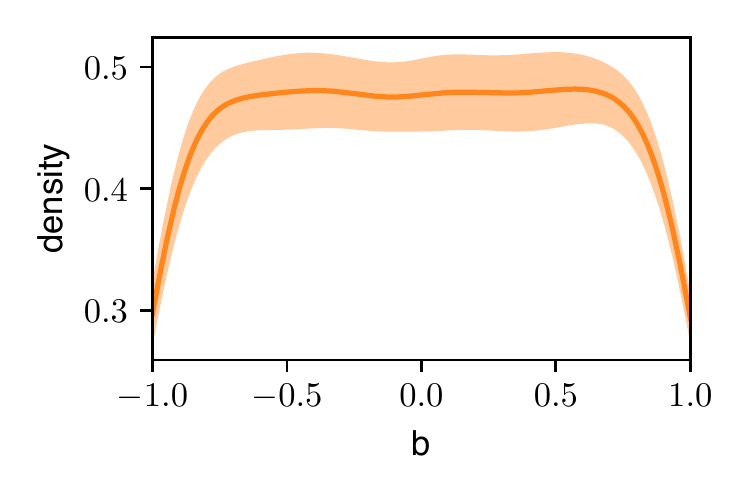}}
    \subfigure[$P_t^{sim}(x)$]{\includegraphics[width=0.24\textwidth]{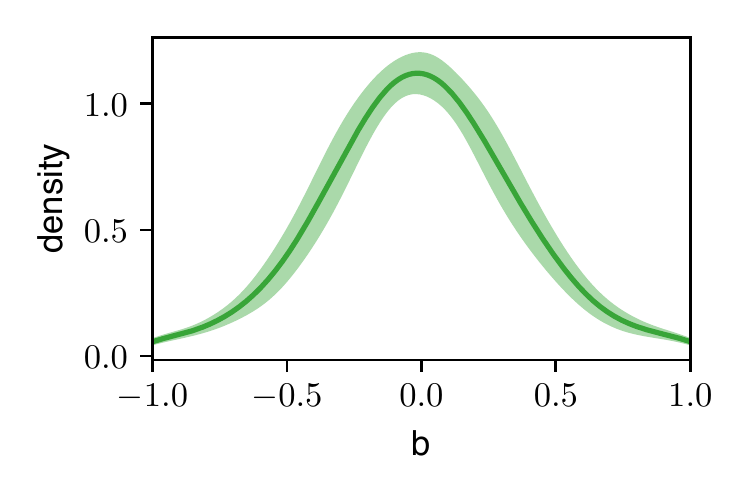}}
    \subfigure[$P_t^{all}(x)$]{\includegraphics[width=0.24\textwidth]{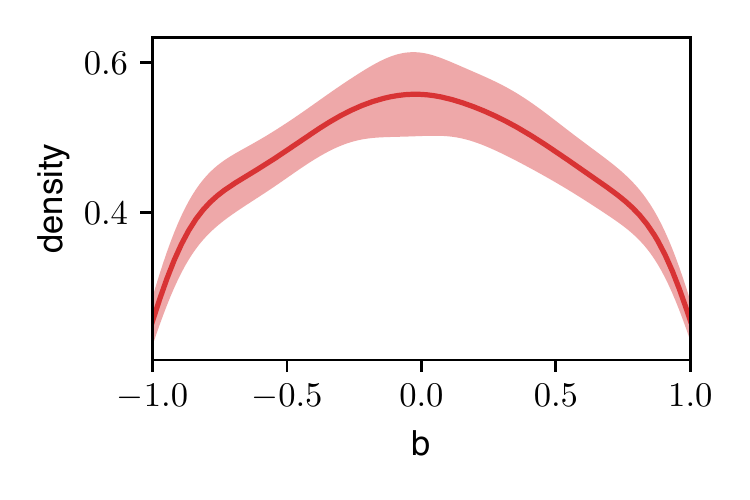}}
    
  \caption{Average opinion distributions obtained from the execution of the dynamics with a fixed reception probability, $P_d^{III}$. Each item represents a different transmission function. The shaded regions show the standard deviations.}
  \label{fig:avgOpEqTransm}
\end{figure*}

We start by comparing the proposed functions of post transmission, where we fixed the distribution function as uniform, $P_d^{III}$, see Figure~\ref{fig:equaltransmission}. The respectively average curves of $b$ are shown in Figure~\ref{fig:avgOpEqTransm}. More specifically, this test represents the cases in which there is no algorithm controlling the social network. The polarized transmission function ($P_t^{\text{pol}}(x)$) give rise to polarized opinions (see Figure~\ref{fig:avgOpEqTransm}(a)). Furthermore, $P_t^{uni}(x)$, results in a uniformly random distribution, where $BC$ is close to $BC_{\text{critc}}$. The border effect affects a little bit the visualization of the density distribution, shown in Figure~\ref{fig:avgOpEqTransm}(b). In the case of individuals that tend to post information similar to their opinions, $P_t^{\text{sim}}(x)$, the consensus is found, see Figure~\ref{fig:avgOpEqTransm}(c). As expected, in the latter case, which considers all functions with the same probability, the resulting opinion distribution seems to be a combination of the previous distributions. This result is reflected in the resulting distribution (Figure~\ref{fig:avgOpEqTransm}(d)) and $BC$ (Figure~\ref{fig:equaltransmission}(d)). In summary, the previous results illustrate that with a fixed reception, the resulting opinion distributions reflect only the transmission function. 

\begin{figure}[!htpb]
  \centering
    \subfigure[$P_d^{I}(x)$]{\includegraphics[width=0.49\textwidth]{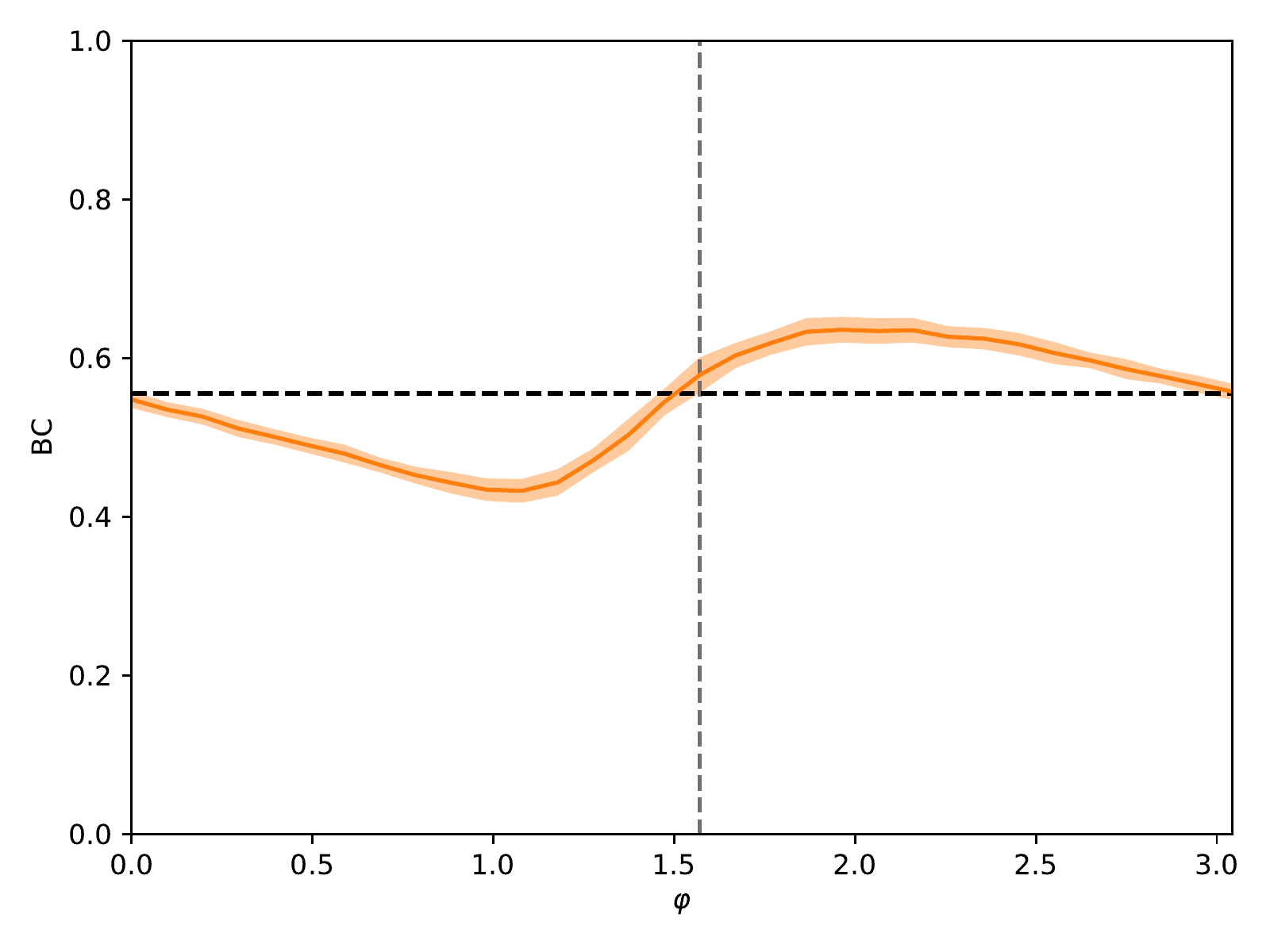}}
    \subfigure[$P_d^{II}(x)$]{\includegraphics[width=0.49\textwidth]{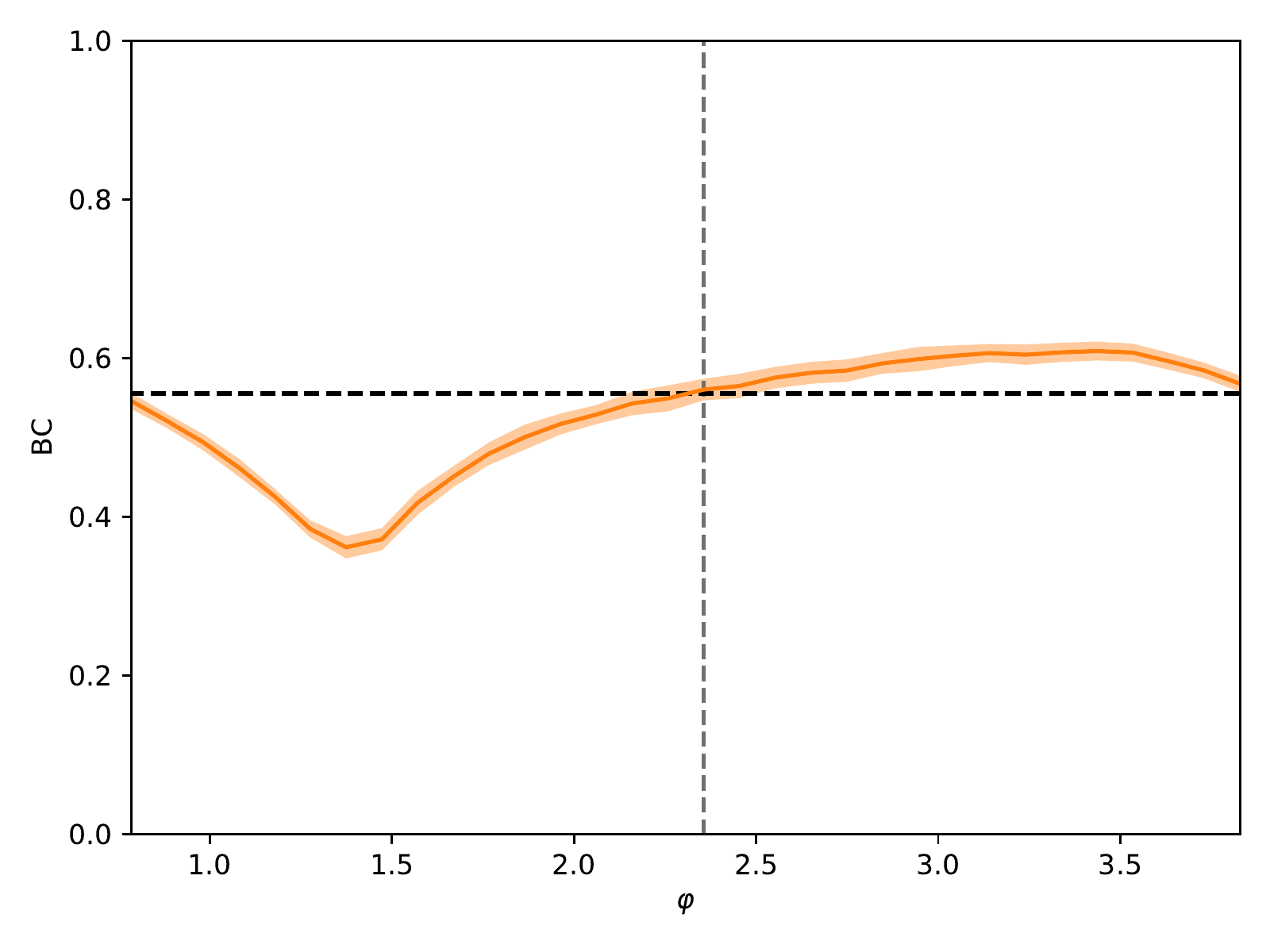}}
    
  \caption{Average $BC$ obtained from the dynamic execution with a fixed transmission probability, $P_t^{\text{uni}}$. Each item represents a different reception function. The shaded regions show the standard deviations, and the horizontal dashed line indicates $BC_{\text{critc}}$. The vertical line in item (a) indicates $\pi/2$ and in item (b) points $3\pi/4$. In the latter, the graph starts and ends at $\pi/4$ and $5 \pi/4$, respectively.}
  \label{fig:equalreception}
\end{figure}

In other to better understand the variations of post distribution, we analyze the results of uniform transmission probability, $P_t^{\text{uni}}$ (see Figure~\ref{fig:equalreception}). In the case of $P_d^{I}(x)$, for $\phi < \pi/2$ and $\phi > \pi/2$ the dynamics tend to converge to uni-modal and bi-modal distributions, respectively. Additionally, the uniform distributions are obtained for $\phi$ values close to 0 or $\pi/2$. By considering $P_d^{II}(x)$, for $\phi < \pi/4$ and $\phi > 3\pi/4$, the dynamics results in bi-modal distributions, otherwise uni-modal, and the regions that tend to uniform distributions are given by $\phi$ close to $\pi/4$ and $3\pi/4$. Due to these differences, in the following of these results, we present the plots with respect to $P_d^{II}(x)$ shifted in $\pi/4$. Note that, since the employed distribution functions are periodic, the results obtained for $\phi = \pi + k$ are equivalent the results obtained for $\phi = k$. 

\begin{figure*}[!htpb]
  \centering
     \subfigure[$P^{I}_d$ and $P_t^{pol}(x)$ ]{\includegraphics[width=0.24\textwidth]{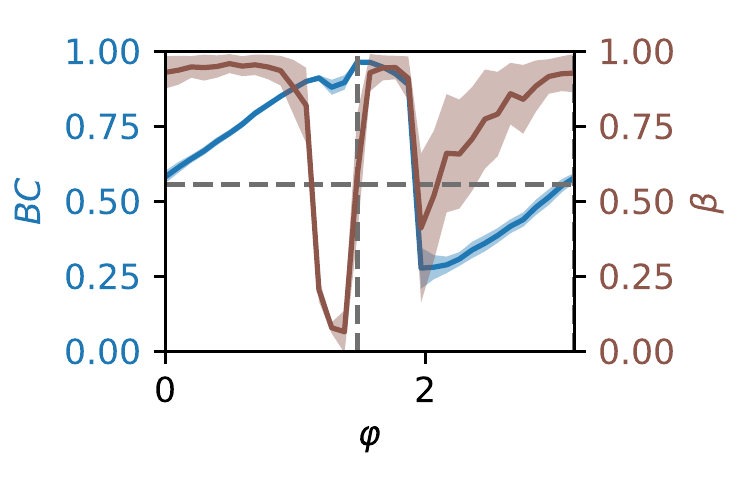}}
    \subfigure[$P^{I}_d$ and $P_t^{uni}(x)$ ]{\includegraphics[width=0.24\textwidth]{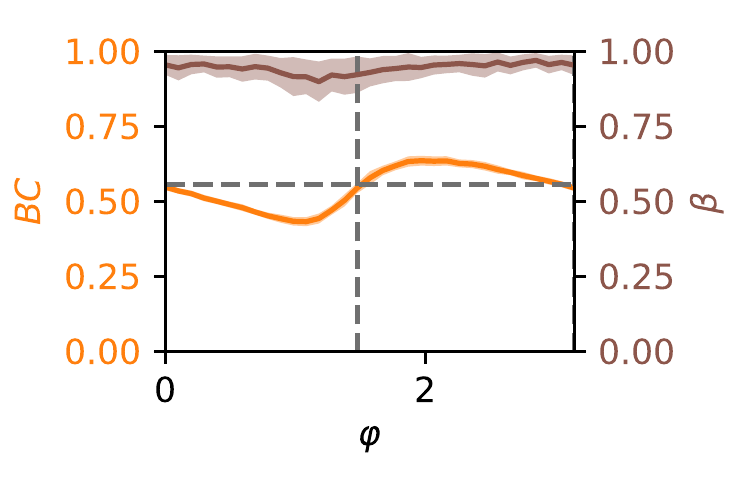}}
    \subfigure[$P^{I}_d$ and $P_t^{sim}(x)$ ]{\includegraphics[width=0.24\textwidth]{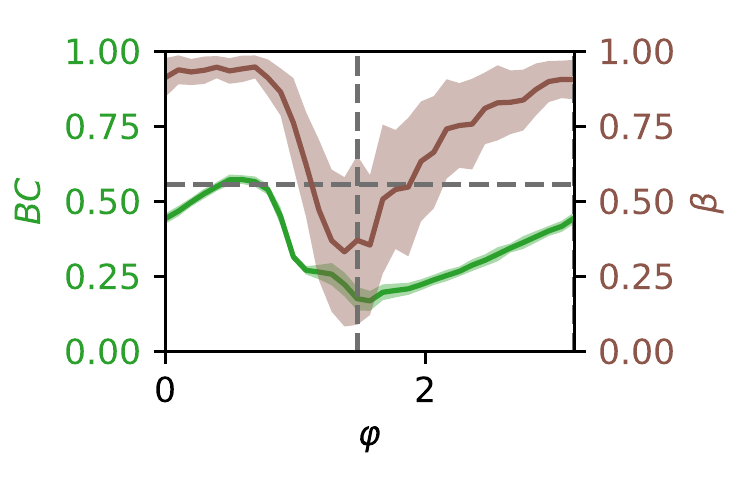}}
    \subfigure[$P^{I}_d$ and $P_t^{all}(x)$ ]{\includegraphics[width=0.24\textwidth]{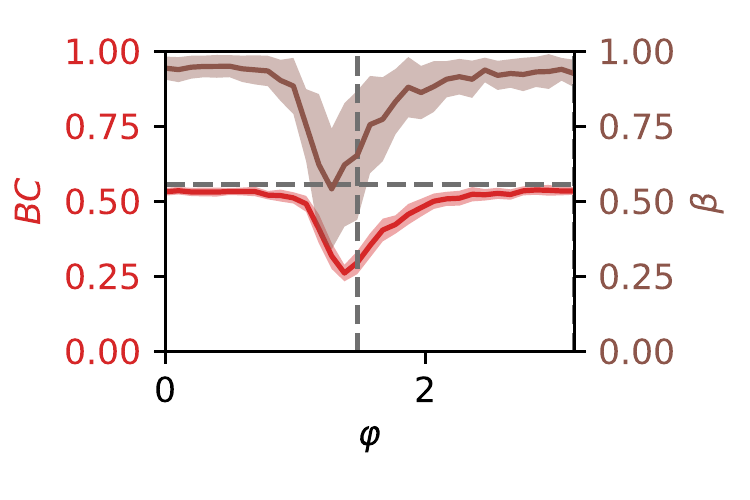}}
    
    \subfigure[$P^{II}_d$ and $P_t^{pol}(x)$ ]{\includegraphics[width=0.24\textwidth]{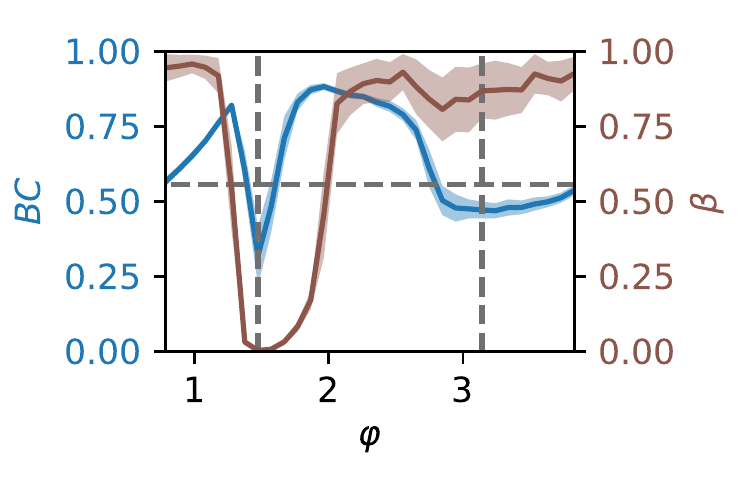}}
    \subfigure[$P^{II}_d$ and $P_t^{uni}(x)$ ]{\includegraphics[width=0.24\textwidth]{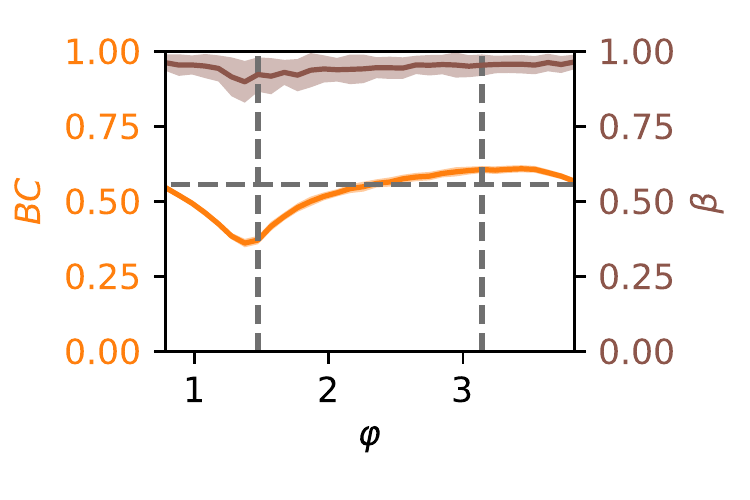}}
    \subfigure[$P^{II}_d$ and $P_t^{sim}(x)$ ]{\includegraphics[width=0.24\textwidth]{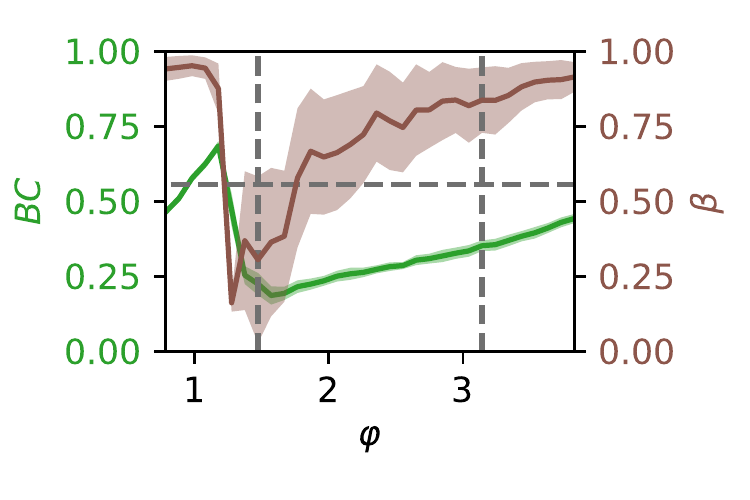}}
    \subfigure[$P^{II}_d$ and $P_t^{all}(x)$ ]{\includegraphics[width=0.24\textwidth]{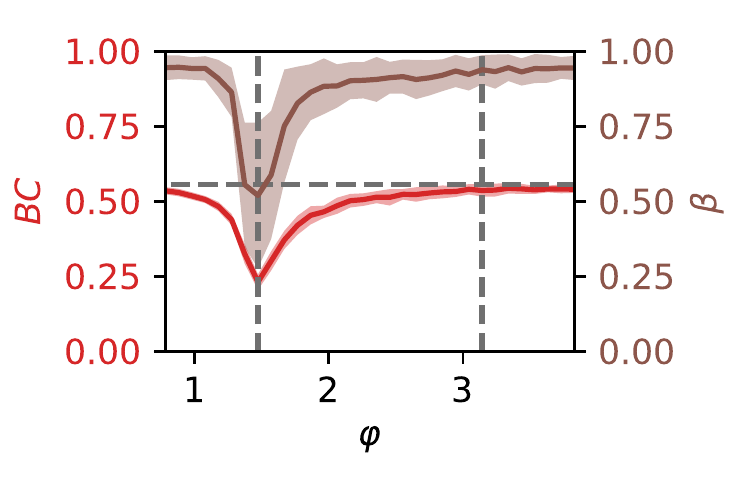}}
    
  \caption{Comparison between $BC$ and $\beta$ (in brown) for all tested parameters. The shaded regions show the standard deviations, and the horizontal dashed line indicates $BC_{\text{critc}}$. The vertical lines in items (a) to (d) indicates $\pi/2$ and in items (e) to (h) indicate $\pi/2$ and $\pi$.}
  \label{fig:bc_x_beta}
\end{figure*}

\section{Comparison between different topologies}
\label{sec:comparison_topologies}
We start this analysis considering the dynamics without the rewiring probability. In order to compare the results, we considered many topologies. The first is the 2D regular lattices with periodic boundary conditions. As another option of networks with a fixed node degree, the \emph{random regular graph} (RRG)~\cite{steger1999generating} was also employed. One important characteristic of natural systems is the power-law degree distribution. So, we incorporate this topology in our analysis through the configuration model~\cite{catanzaro2005generation} with the power parameter equals to $2.2$, with the cutoff $k_{max} = 40$ and $k_{min} = 4$ and $k_{min} = 2$ to have approximately average degrees of 4 and 8, respectively. For this model, we considered the same implementation of~\cite{peron2019onset}. Another feature that can be found in social networks is the presence of communities. Among the many possibilities of community-based networks, we choose the LFR-Benchmark~\cite{lancichinetti2008benchmark}, which also have power-law degree distribution. In this case, we generated networks only with two communities of approximately the same size, and the mixture between the communities was set as $\mu=0.1$. We set the parameters for all of the network models to generate networks with 1000 nodes and average degrees of 4 and 8 approximately. Furthermore, we considered only the biggest connected component.

\begin{figure*}[!htpb]
  \centering
    \subfigure[$P_d^I$ and $\phi=0.0$]{\includegraphics[width=0.48\textwidth]{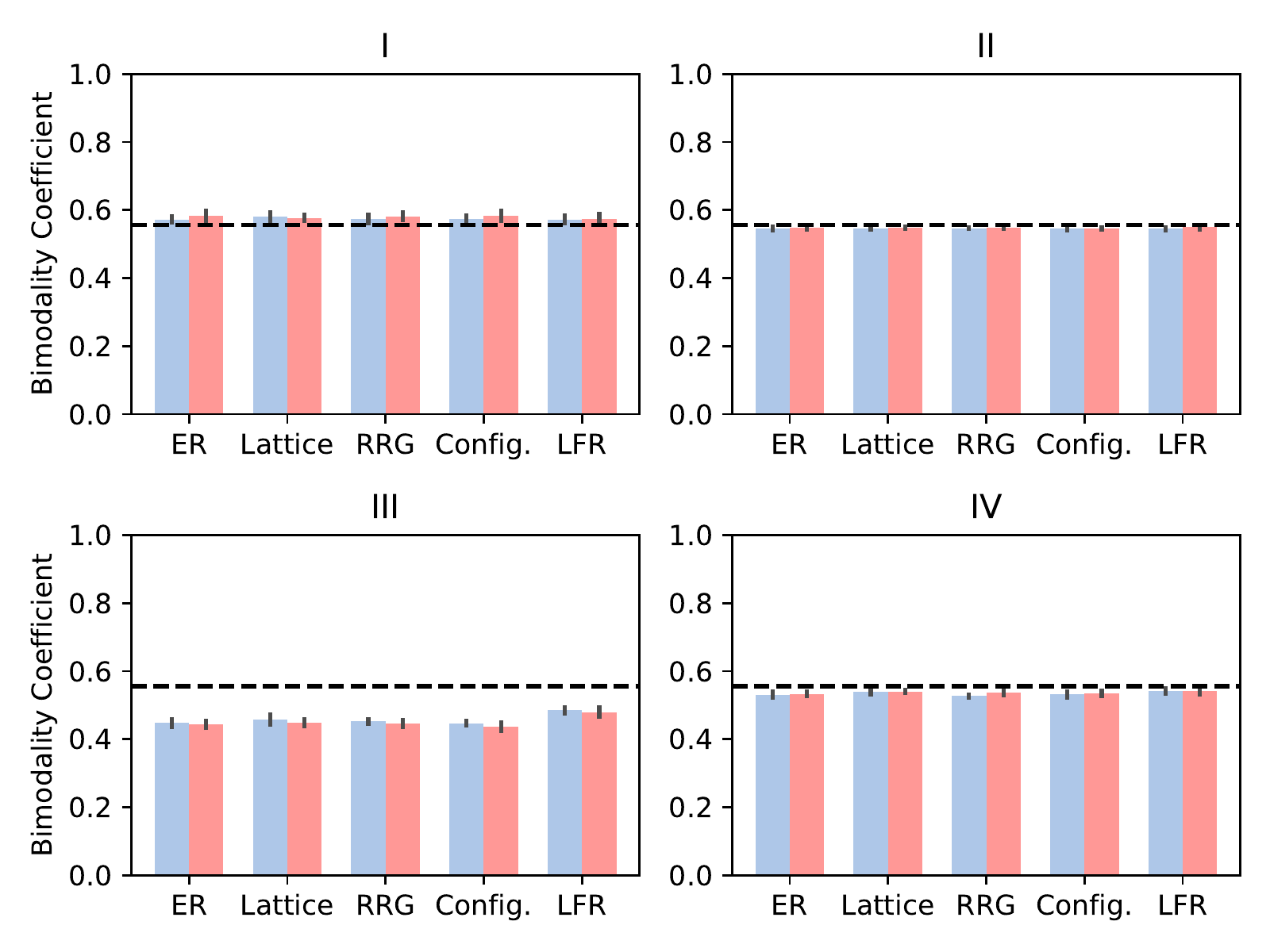}}
    \subfigure[$P_d^{II}$ and $\phi=0.0$.]{\includegraphics[width=0.48\textwidth]{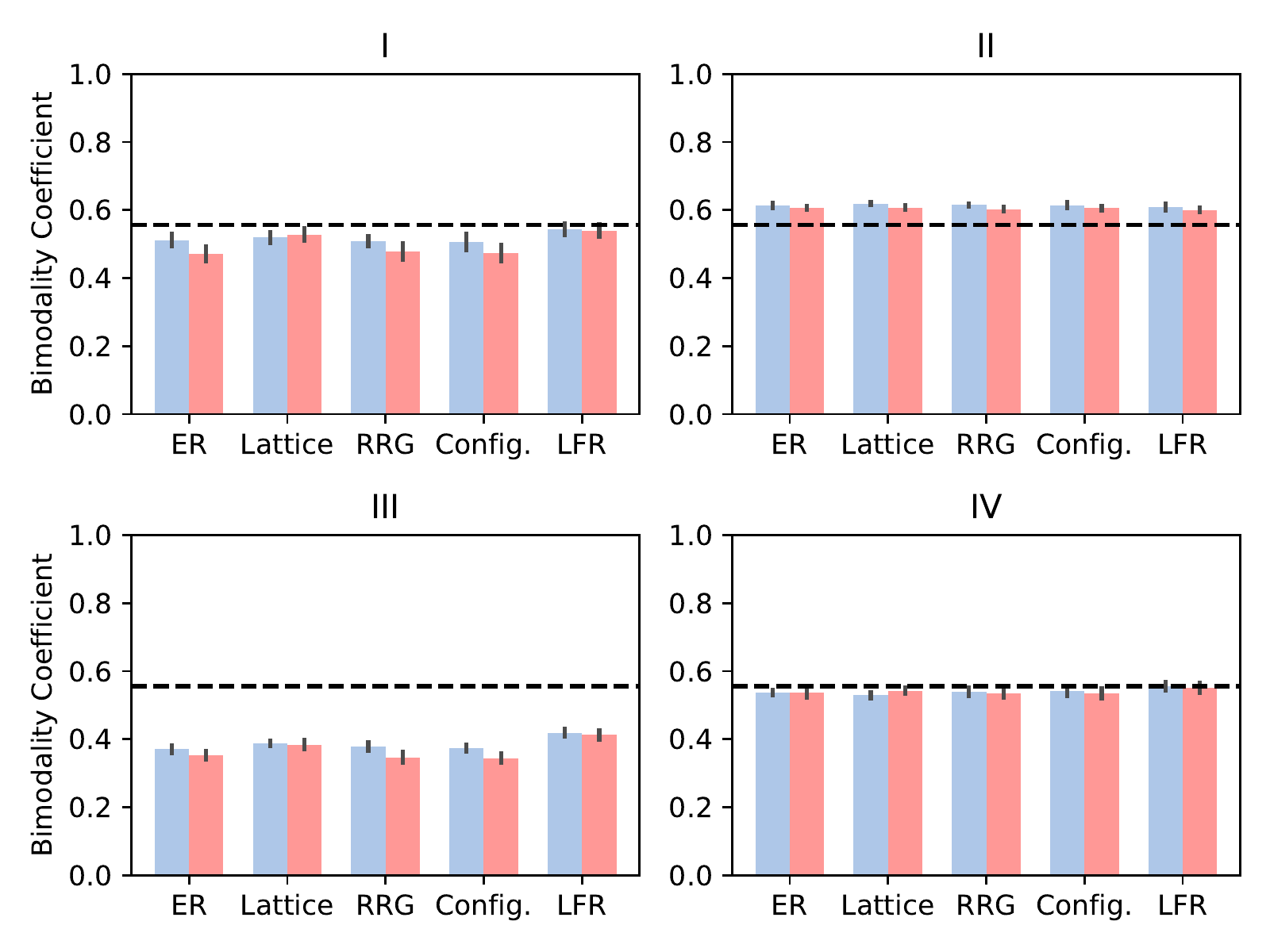}}

    \subfigure[$P_d^I$ and $\phi=1.473$.]{\includegraphics[width=0.49\textwidth]{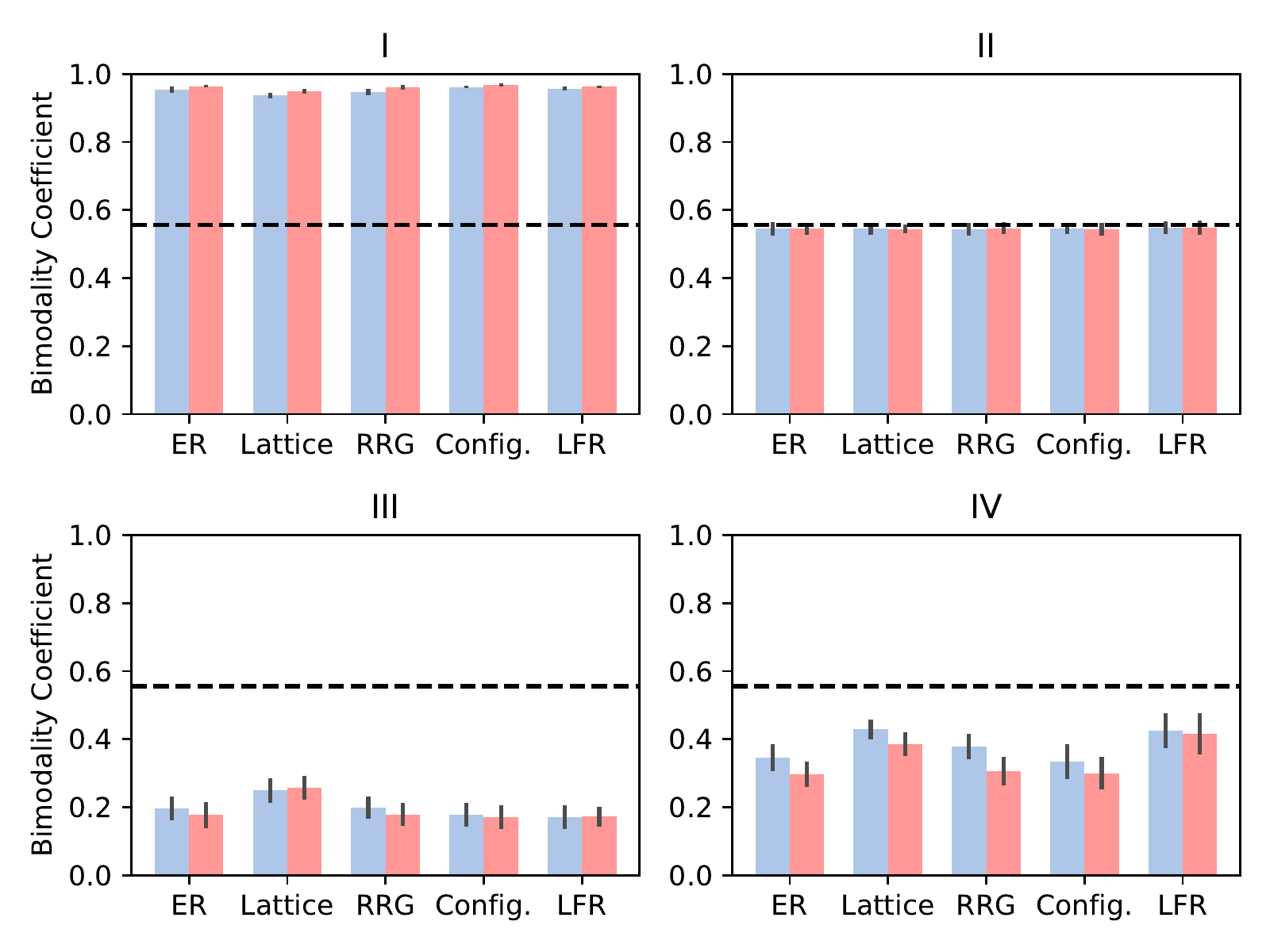}}
    \subfigure[$P_d^{II}$ and $\phi=1.473$]{\includegraphics[width=0.49\textwidth]{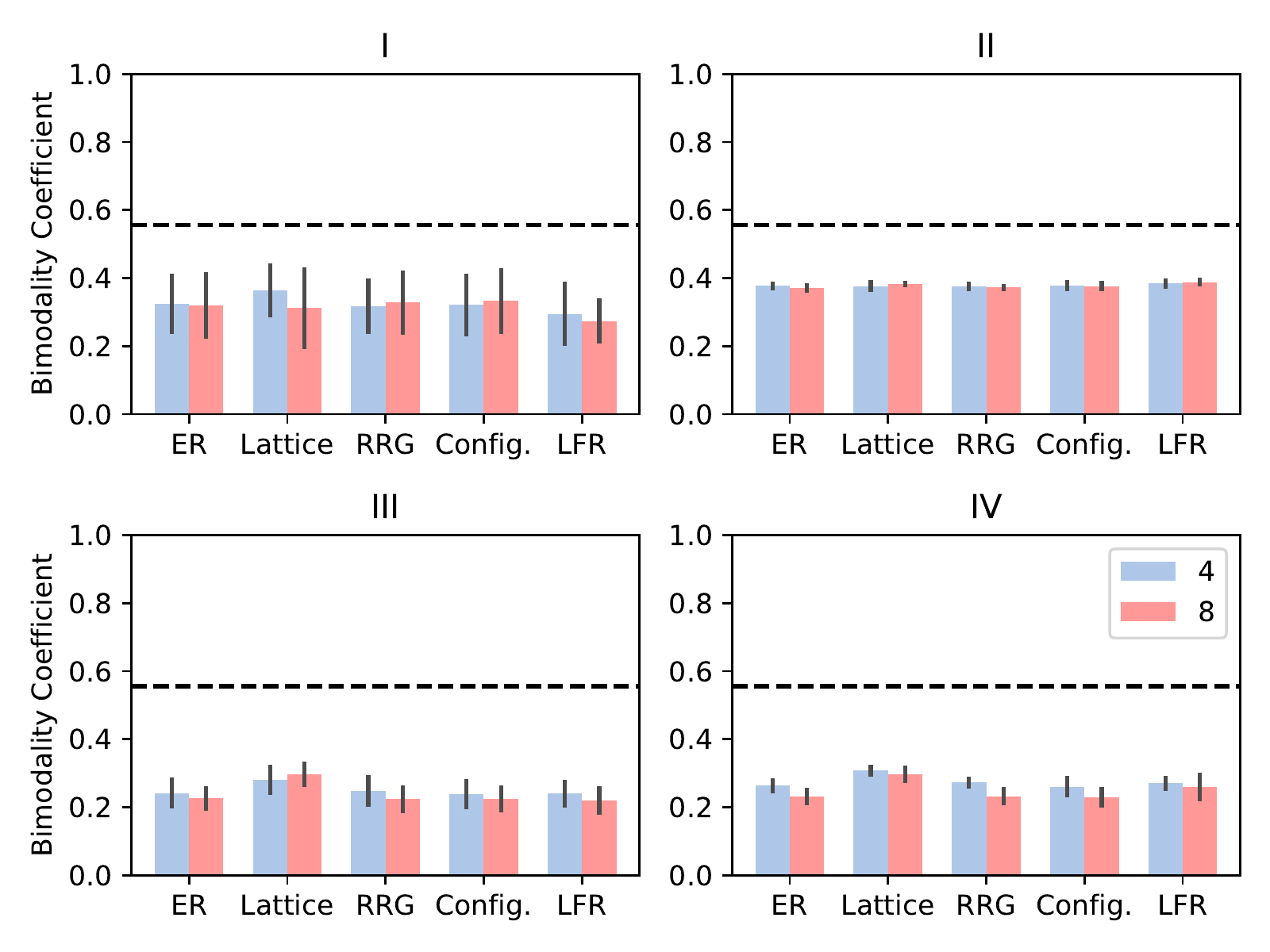}}
    
  \caption{Average bimodality coefficient ($BC$) computed for the selected values of $\phi$ ($0.0$ and $1.473$). The error bars represent the standard deviations, and the colors represent the approximated network average degrees. For each of the sub-figure, we employed the transmission probability functions, as follows: I- $P_t^{pol}(x)$, II- $P_t^{uni}(x)$, III- $P_t^{sim}(x)$, and IV- all functions with the same probability. For all of the cases, there are no significant variations between the results obtained from distinct topologies.}
  \label{fig:comparison}
\end{figure*}

Figure~\ref{fig:comparison} shows the comparison of $BC$ among the network topologies. All in all, $BC$ tends to be similar for all of the compared network topologies. The highest differences are found for $P_t^{pol}$ with $\phi=1.473$, in which $BC$ measured from lattice is smoother than from the remaining network models. Furthermore, the highest standard deviations are found for this parameter combination.

\begin{figure}[!htpb]
  \centering
    \subfigure[Lattice]{\includegraphics[width=0.22\textwidth]{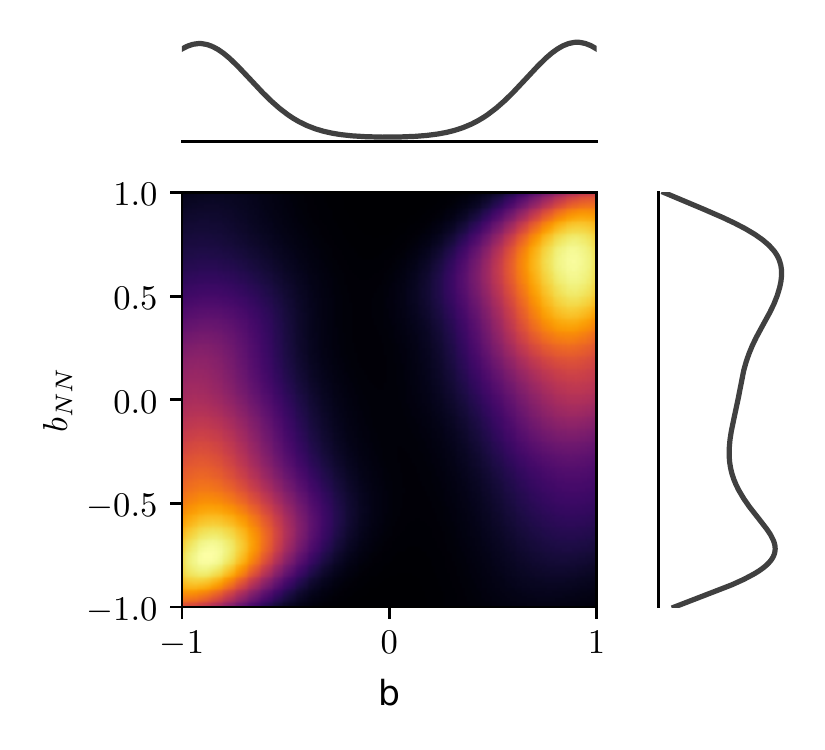}}
    \subfigure[LFR-Benchmark]{\includegraphics[width=0.22\textwidth]{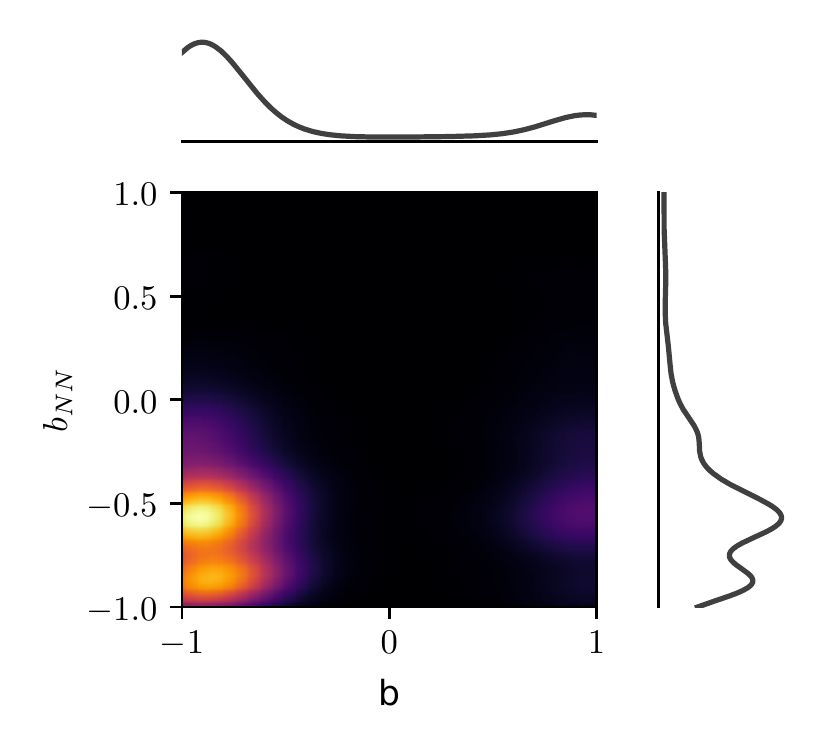}}

  \caption{Density maps measured from the dynamics execution on Lattice and LFR-Benchmark (both with $\langle k \rangle \approx 8$), with the following parameters: $P_t^{pol}$ and $P_d^{I}$ for $\phi=1.473$. Darker colors represent the denser regions.}
  \label{fig:echoChamber2}
\end{figure}

In contrast with most of the topologies, for LFR-Benchmark and Lattice, some different results were found. Figure~\ref{fig:echoChamber2} shows one example of density map obtained from each network topology for $P_t^{pol}$ and $P_d^{I}$ with $\phi=1.473$. Despite the high values of $BC$ (shown in Figure~\ref{fig:comparison}(c)), for both structures, the polarity found in the opinion distributions is not balanced. In the case of Lattice, this topology gave rise to weakly defined echo chambers (see Figure~\ref{fig:echoChamber2}(a)). The highest topological distances between nodes could enable this effect. By considering LFR-Benchmark, a non-trivial result was found, where for many executions of the dynamics, two peaks are found to be close in the density maps (see Figure~\ref{fig:echoChamber2}(b)).

\end{document}